# Time-dependent Photospheric Radiative Transfer in Structured GRB Jets: Spectral Evolution and Polarization Diagnostics

Yue Xu[1,2], Ming Jin[3], and Qingwen Tang[1]
[1] Department of Physics, School of Physics and Materials Science, Nanchang University, Nanchang 330031, People's Republic of China; qwtang@ncu.edu.cn
[2] Henan Institute for Drug and Medical Device Control (Henan Vaccine Issuance Center), Zhengzhou 450008, People's Republic of China
[3] Shanghai Johnson & Johnson Pharmaceuticals, Ltd., Shanghai, People's Republic of China

## Abstract

Photospheric emission from relativistic gamma-ray burst (GRB) jets is a promising mechanism for producing the Band-like spectra observed in the prompt phase, yet the connections between jet structure, dissipation location, and polarization signatures remain unclear. We investigate time-dependent photospheric radiation transfer in structured relativistic jets by coupling two-dimensional axisymmetric special relativistic hydrodynamic simulations with Monte Carlo photon propagation. Photon escape and subphotospheric dissipation are characterized using the residual line-of-sight optical depth $\tau_{\rm out}(\hat{\Omega})$ evaluated along each photon trajectory, allowing a direction-dependent treatment of photon decoupling in structured jets. The radiative transfer includes Klein–Nishina Compton scattering and polarization evolution using the Mueller matrix formalism. We perform a systematic parameter study exploring the effects of viewing angle, electron–positron pair loading ($Z_{\pm}$), and the optical-depth window of subphotospheric dissipation. The model produces time-resolved spectra, peak-energy evolution $E_{\rm pk}(t)$, Band parameters, polarization degree $\Pi(E, t)$, and last-scattering statistics. We find that jet angular structure and the geometry of the line-of-sight optical depth strongly regulate spectral evolution and polarization signatures. The dissipation depth and pair loading jointly control the stability of $E_{\rm pk}$, the formation of high-energy spectral tails, and the energy dependence of polarization. These results provide quantitative predictions for GRB prompt-emission spectra and polarization that can be tested with current and upcoming high-energy polarimeters.

*Unified Astronomy Thesaurus concepts:* Gamma-ray bursts (629); High energy astrophysics (739)

## 1. Introduction

Gamma-ray bursts (GRBs) are among the most luminous transient phenomena in the Universe. During the prompt-emission phase, typically observed in the ∼10 keV–10 MeV band, the spectra are commonly described by the empirical Band function and exhibit pronounced temporal evolution (D. Band et al. 1993; Y. Kaneko et al. 2006; M. Axelsson & L. Borgonovo 2015). In particular, the peak energy $E_{\rm pk}$ often evolves in correlation with the flux, while the low-energy spectral index $\alpha$ and the high-energy tail vary across individual pulses. These time-resolved spectral properties provide important constraints on the radiation mechanisms operating in relativistic GRB jets (T. Piran 2004; P. Mészáros 2006).

In addition to spectral diagnostics, polarization measurements offer an independent probe of jet geometry and radiation physics. The polarization degree $\Pi$ and polarization angle $\chi$ are sensitive to the anisotropy of the radiation field, the geometry of the visible emitting region, and the underlying emission or scattering processes. Consequently, polarization can help distinguish between competing prompt-emission scenarios that produce similar spectral signatures. Recent observations with POLAR and AstroSat/CZTI have reported statistically significant linear polarization in several GRBs (T. Chattopadhyay et al. 2019, 2022; M. Kole et al. 2020), motivating theoretical studies that simultaneously address spectral formation and polarization in relativistic outflows.

### 1.1. Photospheric Emission in Structured Jets

If radiation is produced at sufficiently high optical depth in an ultrarelativistic outflow, photons undergo multiple scatterings before escaping, naturally leading to photospheric emission (J. Goodman 1986; B. Paczynski 1986; D. Giannios 2006). In the simplest fireball picture, radiation and plasma approach thermal equilibrium at $\tau \gg 1$, producing quasi-thermal spectra near a blackbody or Wien peak. However, in realistic GRB jets the photon-decoupling process is intrinsically stochastic and does not occur at a sharply defined geometric radius. Instead, the photosphere corresponds to a statistical last-scattering region whose spatial distribution depends on the local velocity field, density structure, and photon-propagation direction (A. Pe'er 2008; A. M. Beloborodov 2011).

Furthermore, GRB jets are unlikely to be angularly uniform. Hydrodynamic simulations and observational modeling suggest that relativistic jets possess substantial angular structure, such as core–sheath configurations or velocity shear with continuous variations in $\Gamma(\theta)$ and $\rho(\theta)$. In such structured outflows, photons emitted from different angular regions contribute differently to the observed signal. Off-axis viewing introduces variations in Doppler boosting, equal-arrival-time surface (EATS) effects, and geometric asymmetry of the visible region, which can jointly influence the observed spectral evolution and polarization signatures (C. Lundman et al. 2013). Photospheric emission in structured jets therefore constitutes a coupled radiative-transfer problem in which spectral formation and polarization both depend sensitively on jet geometry and scattering physics.

### 1.2. Monte Carlo Radiative Transfer in Relativistic Jets

Theoretical studies of GRB photospheric emission generally follow two complementary approaches. One approach employs radiation hydrodynamics or moment-based formalisms that evolve angular moments of the radiative-transfer equation together with the fluid dynamics. Such treatments provide valuable insight into the global evolution of radiation and plasma under quasi-steady conditions (A. M. Beloborodov 2010; I. Vurm & A. M. Beloborodov 2016) but become increasingly complex when strong angular inhomogeneity, nonlocal scattering, or polarization effects are included.

An alternative approach is to perform photon Monte Carlo simulations on top of a prescribed hydrodynamic background. By explicitly tracking individual photon trajectories and scattering histories, this method naturally captures direction-dependent transport, nonlocal scattering, and polarization evolution, while directly producing observable quantities such as time-resolved spectra and polarization signals (H. Ito et al. 2013; R. Santana et al. 2014; T. M. Parsotan & D. Lazzati 2023). The main limitation of this approach is that radiative feedback on the fluid dynamics is neglected.

In this work, we adopt the Monte Carlo approach and extend it to treat time-dependent radiative transfer in structured relativistic jets. We first perform two-dimensional axisymmetric special relativistic hydrodynamic (SRHD) simulations to obtain time-resolved jet backgrounds. Photons are then propagated through successive simulation snapshots using a piecewise-constant-in-time approximation, allowing photon-propagation delays to accumulate and enabling the construction of time-resolved spectra and polarization signals.

A key aspect of our method is the treatment of photon decoupling in structured jets. Rather than defining the photosphere using a purely radial optical-depth coordinate, we compute the residual optical depth along the actual photon-propagation direction, $\tau_{\rm out}(\hat{\Omega})$, defined as the remaining Thomson optical depth integrated along the photon trajectory toward the observer. We adopt this quantity as the fundamental coordinate for photon escape and subphotospheric dissipation. This coordinate naturally incorporates geometric effects arising from jet structure and viewing angle and provides a unified description of the decoupling process in anisotropic outflows (A. Pe'er 2008; A. M. Beloborodov 2011; C. Lundman et al. 2013; H. Ito et al. 2013).

Within this framework, we present a time-dependent Monte Carlo radiative-transfer model for photospheric emission in structured relativistic jets and investigate its observable implications. Our main results can be summarized as follows: (i) We show that a structured SRHD jet background without additional dissipation or pair loading naturally produces quasi-thermal spectra with limited spectral broadening, providing a baseline for interpreting GRB prompt-emission spectra. (ii) By introducing a parameterized subphotospheric dissipation window ($\tau_{\rm diss,min}$, $\tau_{\rm diss,max}$), we demonstrate that the depth of dissipation strongly regulates the formation of high-energy spectral tails and the evolution of the peak energy $E_{\rm pk}$. (iii) We quantify how electron–positron pair loading, parameterized by $Z_{\pm}$, modifies the effective optical depth, spectral peak energy, and polarization signatures. (iv) We show that the viewing angle $\theta_{\rm obs}$ significantly affects light-curve morphology, time-resolved $E_{\rm pk}(t)$ evolution, and polarization observables owing to geometric and Doppler effects in structured jets. (v) We examine the sensitivity of the results to numerical factors such as snapshot cadence, escape criteria, and spatial resolution, providing guidance for interpreting Monte Carlo simulations of photospheric emission.

Recent multidimensional and Monte Carlo studies have already demonstrated that photospheric emission in structured relativistic outflows can produce rich spectral and polarization behavior (e.g., H. Ito et al. 2013; C. Lundman et al. 2014; T. Parsotan & D. Lazzati 2021a, 2021b). The present work is complementary to these efforts but differs in several respects. First, we explicitly formulate photon escape, dissipation-window selection, and last-scattering statistics using the same direction-dependent residual optical-depth coordinate, $\tau_{\rm out}(\hat{\Omega})$. Second, we emphasize a systematic comparison between the full dynamic treatment and an axis-based reference treatment in order to identify geometry-robust versus geometry-sensitive observables. Third, we combine this comparison with scans in dissipation depth, pair loading, and viewing angle and interpret the resulting differences using the statistics of the extended last-scattering region.

The paper is organized as follows: Sections 2–4 describe the physical framework, the SRHD background, and the Monte Carlo radiative-transfer implementation. Section 5 presents the parameter study, and Section 6 summarizes the implications and limitations of the model.

## 2. Photospheric Model and Numerical Radiative Transfer

In this section we summarize the multizone physical picture of dissipative photospheres (Planck/Wien/unsaturated Comptonization zones) and the role of electron–positron pair loading in regulating the effective optical depth, thereby motivating the expected behavior of the peak energy $E_{\rm pk}$, spectral broadening, and polarization signatures. We then clarify the positioning of our numerical approach relative to analytic radiation-hydrodynamic treatments: rather than solving moment equations with closure relations or performing fully self-consistent radiation–fluid coupling, we post-process time-resolved outputs from a two-dimensional axisymmetric SRHD jet simulation using a photon-by-photon Monte Carlo radiative-transfer scheme. In this framework, the residual line-of-sight optical depth defined along the true propagation direction, $\tau_{\rm out}(\hat{\Omega})$, is adopted as a unified criterion for photon decoupling and as the coordinate that parameterizes subphotospheric dissipation, enabling a consistent treatment of direction-dependent optical-depth effects in structured jets under off-axis geometries. Temporally, we employ a time-dependent treatment in which the background is approximated as piecewise frozen, allowing photons to propagate and scatter across multiple SRHD snapshots while accumulating light-travel time delays, so as to produce time-resolved spectral and polarization diagnostics.

### 2.1. Multizone Photospheric Structure and Subphotospheric Dissipation

In dissipative photospheric models, an outflow expands from optically thick regions to transparency. The multizone structure of the dissipative photosphere and the corresponding photon decoupling geometry are illustrated schematically in Figure 1. The coupling between radiation and plasma varies continuously with radius, but different physical processes dominate in different regimes (B. Paczynski 1986; J. Goodman 1986; A. Pe'er 2008; A. M. Beloborodov 2010; I. Vurm & A. M. Beloborodov 2016):

1. *Planck zone ($\tau_T \gtrsim 10^5$).* Double-Compton emission and free–free emission are efficient, so both photon number and energy approach local thermal equilibrium, yielding a nearly blackbody spectrum. The outflow evolution in this regime is well approximated as adiabatic expansion.
2. *Wien zone ($10^2 \lesssim \tau_T \lesssim 10^5$).* The Compton $y$-parameter satisfies $y \gg 1$, maintaining Compton equilibrium between radiation and electrons. The spectrum approaches a narrow Wien shape (allowing a nonzero chemical potential). The photon number effectively freezes out near the so-called Wien radius $R_W$, which largely sets the eventual peak energy $E_{pk}$.
3. *Unsaturated Comptonization zone ($1 \lesssim \tau_T \lesssim 10^2$).* Subphotospheric dissipation continuously heats electrons (including $e^{\pm}$ pairs), reducing the Compton $y$-parameter to $\mathcal{O}(1)$. A nonthermal high-energy tail develops, while repeated scatterings shape the low-energy spectrum toward a Band-like slope with $\alpha \sim -1$ (A. M. Beloborodov 2010; I. Vurm & A. M. Beloborodov 2016).

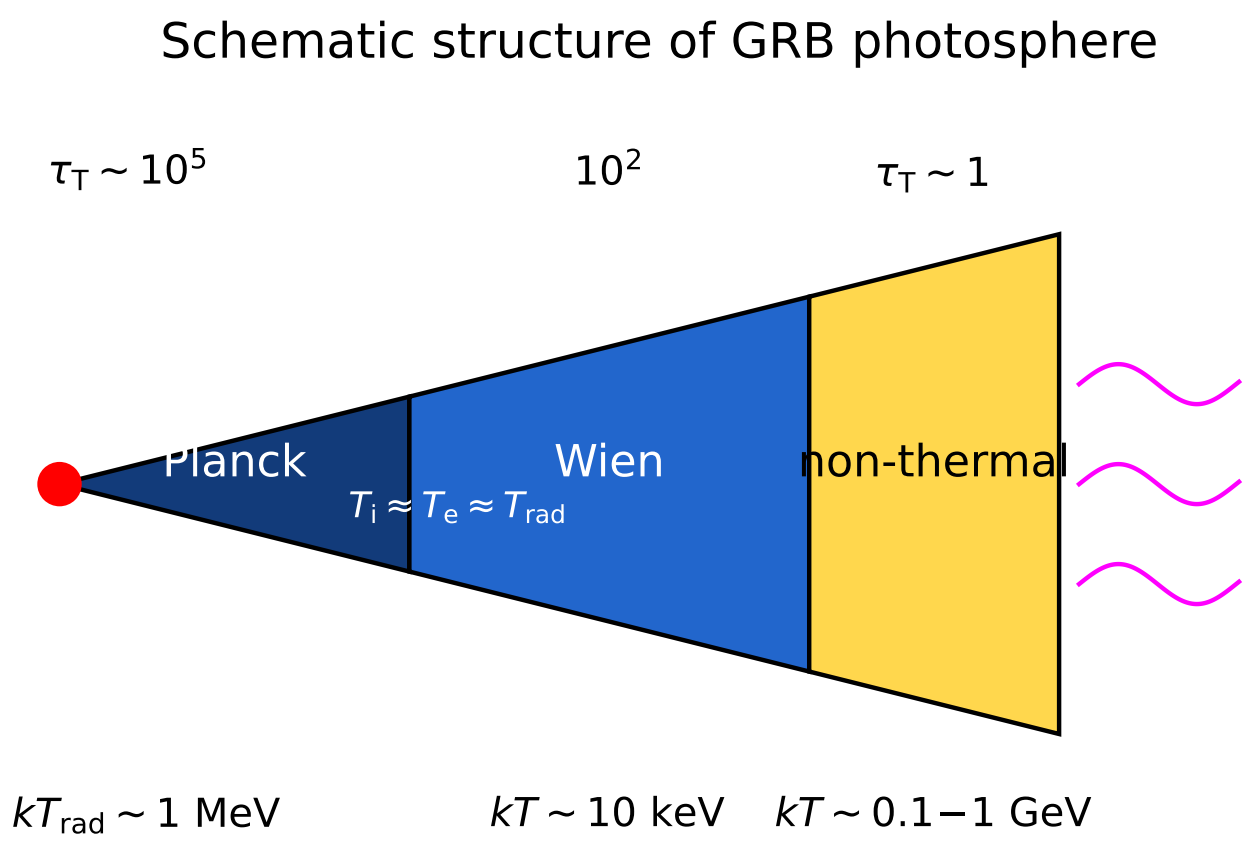

**Figure 1.** Schematic illustration of a multizone dissipative photosphere and the concept of the line-of-sight optical-depth coordinate. The horizontal axis denotes radius $r$, indicating the Planck, Wien, and unsaturated Comptonization zones. In structured jets, photon decoupling is more naturally characterized by the residual optical depth along the true propagation direction, $\tau_{out}(\hat{\Omega})$, rather than by a single radial optical-depth profile $\tau(r)$.

The peak energy $E_{pk}$ is primarily controlled by the ratio of radiation energy density to photon number set in the Planck/Wien zones, whereas spectral broadening, the formation of nonthermal tails, and the evolution of the low-energy slope mainly occur in the unsaturated Comptonization regime at $\tau_T \sim 1$–$10^2$. Because the last-scattering radius typically spans a broad range, $\sim 0.3R_{ph}$–$3R_{ph}$, the photosphere is not a geometrically sharp surface but an extended decoupling region in a statistical sense (A. Pe'er 2008; A. M. Beloborodov 2011). In structured jets viewed off-axis, the shape of this decoupling region depends on the viewing direction, thereby imprinting measurable effects on both spectral and polarization properties.

### *2.2. Electron–Positron Pair Production and Optical-depth Self-regulation*

Within the subphotospheric dissipation region, photon–photon collisions can trigger $\gamma + \gamma \to e^+ + e^-$, substantially increasing the local lepton density. This process has been recognized since early GRB theory as an important ingredient of dissipative outflows (M. J. Rees & P. Mészáros 1994; C. Thompson 1994; P. Mészáros & M. J. Rees 2000). It is convenient to introduce a pair-loading factor, $Z_{\pm} \equiv n_{\pm}/n_p$, to quantify the enhancement of the lepton number density relative to protons.

Numerical studies indicate that $Z_{\pm}$ can reach $\sim 10^2$ in radiation-mediated shock fronts and then decrease downstream owing to pair annihilation (A. Levinson & E. Nakar 2020). The MeV photons required for pair creation may arise from bulk Comptonization associated with velocity shear and may also be produced by inverse Compton scattering of hot electrons if collisionless substructures develop within the shock transition.

Sustained $e^{\pm}$ production can strongly modify the effective optical depth. Even if the background electron–proton plasma would otherwise become transparent, a dissipative layer may remain at $\tau_T \gtrsim 1$ owing to pair loading, keeping radiation coupled to the plasma out to larger radii (H. Ito et al. 2013; A. M. Beloborodov 2017). In this case the photosphere is no longer associated with a fixed radius; rather, it is effectively "carried" outward by the pair-supported dissipative layer as the jet expands.

We stress that, in principle, the spatial distribution of $Z_{\pm}$ should be determined self-consistently by the radiation field together with pair creation and annihilation microphysics. Here our aim is to quantify the sensitivity of observable diagnostics to pair loading in a parameterized manner. We therefore treat $Z_{\pm}$ as an external control parameter and scan its values in numerical experiments. Its primary impact is to regulate the effective optical depth and the geometry of the last-scattering surface, thereby shaping the responses of $E_{pk}$, spectral form, and polarization to pair loading.

### *2.3. Line-of-sight Optical Depth in Structured Jets*

In spherically symmetric, steady outflows, the radial optical depth $\tau(r)$ is often used as an approximate "photospheric coordinate," from which a photospheric radius $R_{ph}$ can be defined. In structured jets, however, both density and Lorentz factor vary with polar angle, and shear layers and boundary structures are present, so the decoupling condition is no longer equivalent in different propagation directions. The physically relevant decoupling depth for a given photon is the residual optical depth along its own propagation direction,

$$\tau_{out}(\hat{\Omega}) = \int_{ray} n_e' \sigma_T \, ds', \qquad ds' = \Gamma(1 - \boldsymbol{\beta} \cdot \hat{\Omega}) \, ds. \quad (1)$$

This definition naturally incorporates nonradial structure and viewing-geometry effects: at the same location, different propagation directions may correspond to substantially different decoupling depths, thereby altering the last-scattering distribution and the efficiency of spectral broadening and polarization generation (A. Pe'er 2008; A. M. Beloborodov 2011; H. Ito et al. 2013; C. Lundman et al. 2014). Accordingly, in our Monte Carlo implementation we evaluate $\tau_{out}(\hat{\Omega})$ dynamically along each photon trajectory and use it consistently as the escape criterion, the coordinate defining dissipation windows, and the reference quantity for last-scattering statistics, ensuring an algorithmically unified treatment of geometric effects in structured jets.

**Table 1**
Summary of the Fiducial SRHD Setup Used to Generate the Structured-jet Backgrounds for the Monte Carlo Radiative-transfer Calculations

| Category | Quantity | Value/Description |
|---|---|---|
| Geometry | Coordinate system | Two-dimensional axisymmetric cylindrical coordinates $(r, z)$ |
| Spacetime | Metric | Flat Minkowski spacetime |
| Hydrodynamics | Dynamical model | SRHD, without magnetic fields |
| Equation of state | EOS | Ideal-gas ($\gamma$-law) EOS with $\gamma_{\rm ad} = 4/3$ |
| Numerical scheme | Spatial discretization | Finite-volume Godunov scheme |
| Numerical scheme | Riemann solver | HLL approximate solver |
| Numerical scheme | Spatial reconstruction | Piecewise-constant (first-order Godunov) reconstruction in the fiducial production runs; optional MUSCL-type piecewise-linear reconstruction with minmod limiter for comparison tests |
| Numerical scheme | Time integration | Two-stage RK2 (Heun) |
| Numerical scheme | CFL number | CFL = 0.3 |
| Grid | Domain size | $R_{\rm max} = 30$ code units, $Z_{\rm max} = 100$ code units |
| Grid | Resolution | $N_r = 256$, $N_z = 768$ |
| Grid | Ghost zones | $n_g = 2$ |
| Units | Length unit | $L_0 = 10^{10}$ cm |
| Units | Time unit | $t_0 = L_0/c \simeq 0.3336$ s |
| Units | Density unit | $\rho_0 = 10^{-7}$ g cm$^{-3}$ |
| Jet injection | Isotropic-equivalent luminosity | $L_{\rm iso} = 10^{52}$ erg s$^{-1}$ |
| Jet injection | Half-opening angle | $\theta_j = 5°$ |
| Jet injection | Initial Lorentz factor | $\Gamma_0 = 50$ |
| Jet injection | Initial specific enthalpy | $h_0 = 20$ |
| Jet-injection geometry | Virtual nozzle height | $z_{0,\rm inj} = 1.0$ code units |
| Jet-injection geometry | Inlet jet radius | $r_j = z_{0,\rm inj} \tan\theta_j$ |
| Jet-injection history | Inflow prescription | Steady nozzle inflow maintained throughout the SRHD run up to $t_{\rm end}$ |
| Ambient medium | Density profile | $\rho_{\rm amb}(R_{\rm sph}) = \rho_{\rm ref}(R_{\rm sph}/r_{\rm ref})^{-k}$, with $R_{\rm sph} = \sqrt{r^2 + z^2}$ |
| Ambient medium | Profile parameters | $k = 2.5$, $\rho_{\rm ref} = 10^{-7}$ g cm$^{-3}$, $r_{\rm ref} = 1.0$ code units |
| Floors | Density floor | $\rho_{\rm floor} = 10^{-10}$ (code units) |
| Floors | Pressure floor | $p_{\rm floor} = 10^{-10}$ (code units) |
| Boundary conditions | Axis boundary | Reflective at $r = 0$ |
| Boundary conditions | Outer boundaries | Outflow at the outer radial and upper $z$ boundaries |
| Boundary conditions | Lower boundary | Nozzle inflow in the jet region; ambient prescription elsewhere |
| Run control | Final time | $t_{\rm end} = 120$ code units |
| Run control | Maximum number of steps | $2 \times 10^5$ |
| Outputs | Snapshot cadence | Every 200 time steps in the fiducial setup |
| Outputs | Stored quantities | $r_{\rm cent}$, $z_{\rm cent}$, $\rho$, $p$, $v_r$, $v_z$, $\Gamma$, $t_k$, and step index |

## 3. SRHD Background: Equations, Numerical Implementation, and Data Output

We perform two-dimensional axisymmetric ($r$–$z$) SRHD simulations to generate the jet background, which serves as the external input for the subsequent Monte Carlo photospheric radiative transfer. In this section we summarize the governing equations, numerical scheme, and output strategy. Further implementation details (primitive-variable recovery, floor treatment, axis handling, etc.) are deferred to the Appendix.

Table 1 summarizes the fiducial SRHD configuration, including the numerical scheme, jet-injection setup, boundary conditions, code-unit normalization, and output strategy.

### 3.1. Governing Equations and Equation of State

The SRHD equations are solved in flat Minkowski spacetime without magnetic fields, adopting units with $c = 1$. The primitive variables are $\{\rho, p, v_r, v_z\}$, where $\rho$ is the comoving rest-mass density. The conserved variables are defined as

$$D = \rho W, \quad S_r = \rho h W^2 v_r, \tag{2}$$

$$S_z = \rho h W^2 v_z, \quad \tau = \rho h W^2 - p - D, \tag{3}$$

where $W = (1 - v^2)^{-1/2}$ is the Lorentz factor and $v^2 = v_r^2 + v_z^2$. The specific enthalpy follows an ideal-gas ($\gamma$-law) equation of state,

$$h(\rho, p) = 1 + \frac{\gamma_{\rm ad}}{\gamma_{\rm ad} - 1}\frac{p}{\rho}, \qquad \gamma_{\rm ad} = 4/3, \tag{4}$$

which yields the sound speed

$$c_s^2 = \frac{\gamma_{\rm ad} p}{\rho h}. \tag{5}$$

These expressions follow the standard conservative formulation of SRHD (J. M. Martí & E. Müller 1996, 1999; E. F. Toro 1999).

In axisymmetric cylindrical coordinates, the conservation equations can be written as

$$\partial_t \boldsymbol{U} + \partial_r \boldsymbol{F}_r(\boldsymbol{U}) + \partial_z \boldsymbol{F}_z(\boldsymbol{U}) = \boldsymbol{S}_{\rm axi}, \tag{6}$$

where $\boldsymbol{U} = (D, S_r, S_z, \tau)^{\mathsf{T}}$. In the present implementation, the cylindrical geometry is handled through a Cartesian-like flux-divergence update supplemented by an explicit axisymmetric geometric source term in the radial momentum equation. In particular, the source contribution proportional to $+p/r$ is added explicitly to the $S_r$ equation. This patch form is adopted for numerical robustness near the symmetry axis, while reflective boundary conditions are imposed at $r = 0$. Additional implementation details, including the numerical handling near the axis, are given in the Appendix (J. M. Martí & E. Müller 1999).

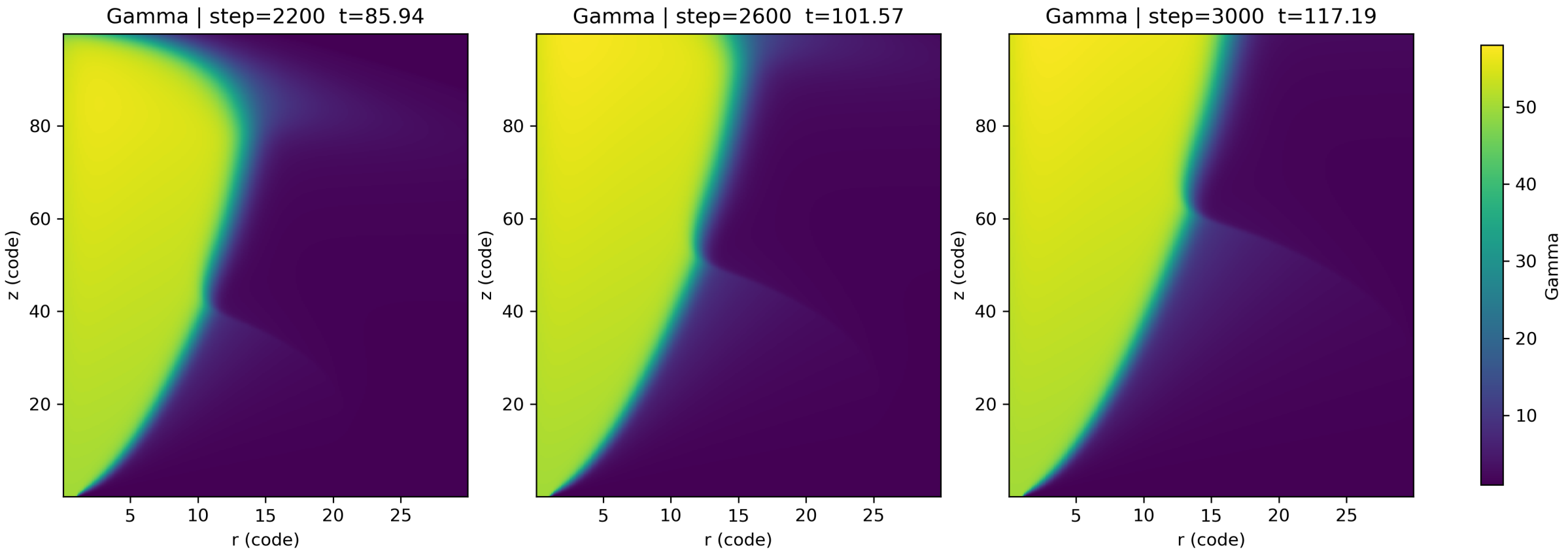

**Figure 2.** Lorentz factor distribution $\Gamma(r, z)$ of the two-dimensional axisymmetric SRHD jet background at representative output times, shown in the $(r, z)$-plane, with $r$ on the horizontal axis and $z$ on the vertical axis. The high-$\Gamma$ jet core extends to large heights, while strong transverse gradients delineate the boundary between the relativistic jet and the surrounding slower material. This structured flow field provides the dynamical background for the subsequent Monte Carlo photospheric radiative-transfer calculations.

### 3.2. Numerical Scheme: Finite Volume, HLL Flux, and RK2 Time Integration

Spatial discretization is performed within a finite-volume Godunov framework. Intercell fluxes are computed using the HLL (Harten–Lax–van Leer) approximate Riemann solver (A. Harten et al. 1983; E. F. Toro 1999; R. J. LeVeque 2002):

$$\boldsymbol{F}_{\rm HLL} = \begin{cases} \boldsymbol{F}(\boldsymbol{U}_L), & s_L \geqslant 0, \\ \boldsymbol{F}(\boldsymbol{U}_R), & s_R \leqslant 0, \\ \dfrac{s_R \boldsymbol{F}(\boldsymbol{U}_L) - s_L \boldsymbol{F}(\boldsymbol{U}_R) + s_L s_R (\boldsymbol{U}_R - \boldsymbol{U}_L)}{s_R - s_L}, & s_L < 0 < s_R, \end{cases} \tag{7}$$

where $s_L$ and $s_R$ denote the minimum and maximum signal speeds estimated from the relativistic sound speed and normal velocities (see the Appendix for details).

The time step is determined by a Courant–Friedrichs–Lewy (CFL) condition,

$$\Delta t = \mathrm{CFL}\, \min\left(\frac{\Delta r}{s_{\max,r}}, \frac{\Delta z}{s_{\max,z}}\right), \tag{8}$$

and time integration is performed using a second-order Runge–Kutta (Heun) scheme,

$$\boldsymbol{U}^{(1)} = \boldsymbol{U}^n + \Delta t\, \mathcal{R}(\boldsymbol{U}^n), \tag{9}$$

$$\boldsymbol{U}^{n+1} = \boldsymbol{U}^n + \frac{\Delta t}{2}[\mathcal{R}(\boldsymbol{U}^n) + \mathcal{R}(\boldsymbol{U}^{(1)})], \tag{10}$$

where $\mathcal{R}$ represents the combined flux divergence and geometric source terms. By default, piecewise-constant reconstruction is adopted in the fiducial production runs for numerical robustness. This choice reflects the role of the SRHD calculation in the present work, namely to provide a stable large-scale structured background for Monte Carlo post-processing rather than to study shock microphysics in detail. An optional second-order MUSCL-type piecewise-linear reconstruction with a minmod limiter is implemented for comparison tests (B. van Leer 1979; E. F. Toro 1999).

### 3.3. Jet Injection, Boundary Conditions, and Ambient Medium

Reflective boundary conditions are imposed at the symmetry axis ($r = 0$), while outflow conditions are applied at the outer radial boundary and the upper $z$ boundary. The lower boundary is treated as a mixed boundary: steady nozzle inflow is imposed inside the jet inlet, while the ambient prescription is applied outside the injected jet region.

The jet is injected through a virtual nozzle geometry with injection radius

$$r_j = z_0 \tan\theta_j, \tag{11}$$

where the injection state is specified by $(\Gamma_0,\ h_0)$ and the isotropic-equivalent luminosity $L_{\rm iso}$. The inlet density and pressure normalization are obtained from energy-flux scaling relations (see the Appendix for details). Such setups are commonly adopted in SRHD simulations of GRB jets (M. A. Aloy et al. 2000; W. Zhang et al. 2003; B. J. Morsony et al. 2007). The ambient medium is modeled as a spherically symmetric power-law density profile.

### 3.4. Output Snapshots and Temporal Sampling

The SRHD background is stored as a sequence of Hierarchical Data Format version 5 (HDF5) files. Each snapshot records $\{r_{\rm cent},\ z_{\rm cent}\}$ and the primitive variables $\rho$, $p$, $v_r$, $v_z$, and $\Gamma$, together with the corresponding physical time $t_k$.

Because the time step varies adaptively according to the CFL condition, outputs written at fixed iteration intervals are not strictly equally spaced in time. We therefore define an effective frame rate,

$$\mathrm{fps}_{\rm eff} = [\mathrm{median}(t_{k+1} - t_k)]^{-1}. \tag{12}$$

### 3.5. Representative Background Structure: Lorentz Factor, Density, and Pressure

Figures 2–4 show representative snapshots of the axisymmetric SRHD jet background at selected output times,

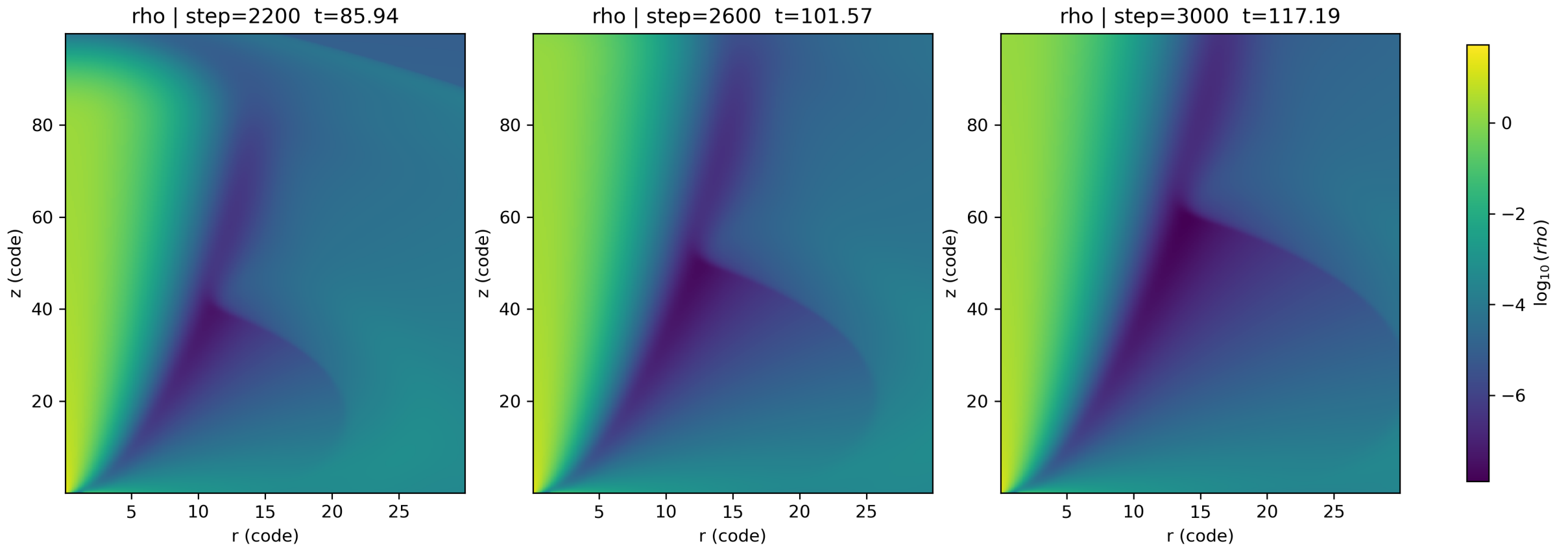

**Figure 3.** Density distribution at representative output times, shown as $\log_{10}\rho(r, z)$ in the $(r, z)$-plane. The strong density contrast between the jet interior, the compressed interface region, and the ambient medium determines the local electron number density and therefore the effective scattering optical depth relevant for photon transport and decoupling.

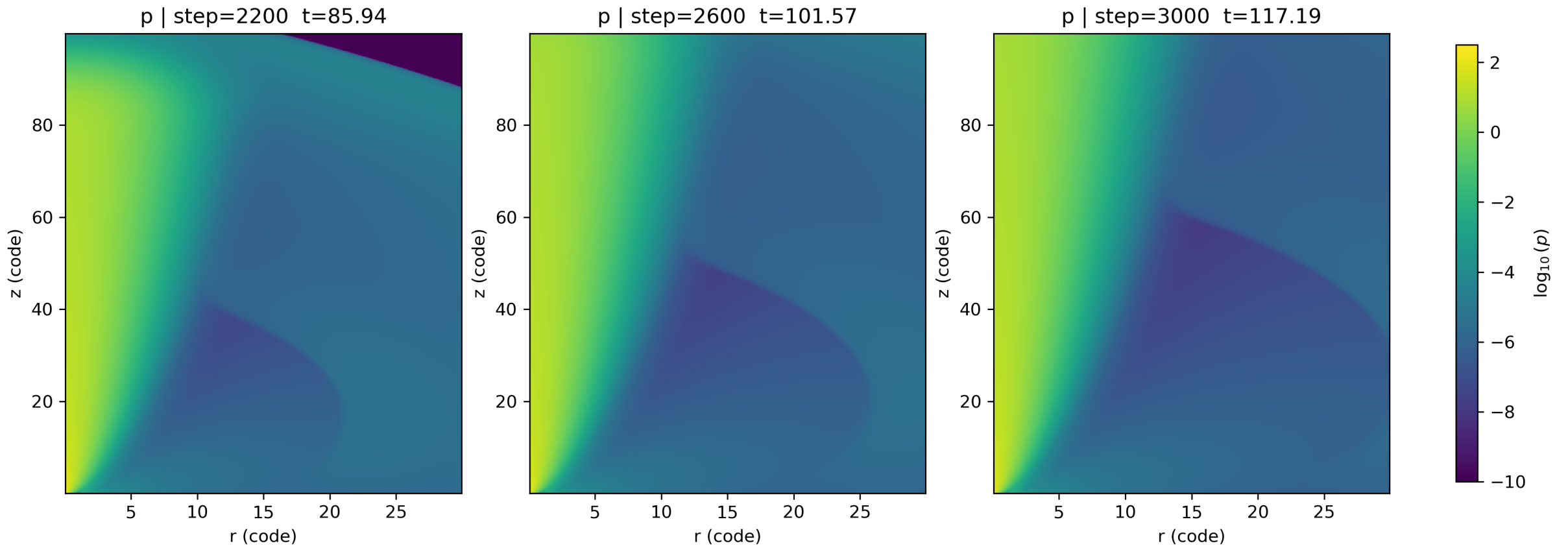

**Figure 4.** Pressure distribution at representative output times, displayed as $\log_{10}p(r, z)$ in the $(r, z)$-plane. The pressure structure traces the thermal content of the jet and the large-scale compression pattern near the jet-head/cocoon region and provides the background thermodynamic scaling for the parameterized heating prescriptions adopted in the radiative-transfer calculations.

displayed in the $(r, z)$-plane with $r$ on the horizontal axis and $z$ on the vertical axis. The Monte Carlo radiative transfer primarily uses snapshots within the interval `collapsar_srhd_step002000-003000.h5`, with the physical time $t_k$ stored in each file and used consistently in post-processing.

The flow exhibits a relativistic jet core extending to large heights, together with pronounced transverse gradients and boundary structure at the interface with the surrounding medium. The Lorentz factor maps show that the high-Γ core remains well collimated while the outer regions are substantially slower, producing a structured-jet–cocoon configuration. The density and pressure maps show corresponding inhomogeneities and compressed regions near the head/interface zone. These spatial variations determine the local electron number density, the effective scattering optical depth, and the angular dependence of photon decoupling in the subsequent Monte Carlo calculations (see Section 4). The presence of resolved compression/interface structure is illustrated more directly by the transverse-cut diagnostics in Appendix A.5, which show sharp radial gradients in $\rho$, $p$, and Γ across the jet-head/cocoon region at representative times.

The role of these figures is to document the large-scale SRHD background that underlies the radiative-transfer calculations. For a more explicit diagnostic of the resolved compression structure, Appendix A.5 provides transverse cuts through the compressed jet-head region, showing the corresponding radial variations of $\log_{10}\rho$, $\log_{10}p$, and Γ.

## 4. Monte Carlo Implementation

In this section, we describe the numerical implementation of the Monte Carlo radiation transfer (MCRT) method adopted in this work (S. Chandrasekhar 1960; G. B. Rybicki & A. P. Lightman 1979; L. A. Pozdnyakov et al. 1983; D. Mihalas & B. W. Mihalas 1984). For clarity and reproducibility, Tables 2 and 3 summarize the fiducial Monte Carlo setup, together with the main parameter scans discussed below.

Because several referee comments concerned the interpretation of code units and observer-frame scales, Table 4 explicitly lists the mapping between the adopted code units and the physical or post-processed observer quantities used throughout the paper.

Unlike analytical treatments based on continuous radiation hydrodynamics (e.g., moment equations of the radiative-transfer equation yielding luminosity evolution, jet acceleration, and pair-loading factor evolution $Z_\pm = n_\pm/n_p$, demonstrating how subphotospheric dissipation sustains optical depth and reshapes the peak energy $E_{\rm pk}$; see A. M. Beloborodov 2010; I. Vurm & A. M. Beloborodov 2016), our numerical approach follows individual photon packets in a prescribed background flow field, neglecting radiative feedback on the hydrodynamics (H. Ito et al. 2013; R. Santana et al. 2014; T. M. Parsotan & D. Lazzati 2023).

**Table 2**
Summary of the Fiducial Monte Carlo Transport and Injection Setup

| Category | Quantity | Value/Description |
|---|---|---|
| Background input | Hydrodynamic source | Time-ordered HDF5 snapshots from the 2D axisymmetric SRHD runs |
| Background treatment | Time dependence | Piecewise constant in time; photons may propagate across multiple SRHD snapshots |
| Background treatment | Frame subsampling | Fiducial: every available frame; comparison runs use every $n$th frame |
| Optical-depth treatment | Dynamic mode | $\tau_{\rm out}(\hat{\Omega})$ evaluated along the actual photon-propagation direction |
| Optical-depth treatment | Axis reference mode | Simplified axis-based optical-depth treatment used as a comparison model |
| Pair loading | Electron density prescription | $n_e' = (1 + Z_\pm)\, \rho/m_p$ |
| Injection geometry | Injection plane | Fixed plane at $z_{\rm init} = 1.0$ code units unless otherwise stated |
| Injection geometry | Injection radius | $R_{\rm inj} = 2.0$ code units |
| Injection verification | Initial optical-depth check | Fiducial requirement $\tau_{\rm init,check} \gtrsim 30$ |
| Injection spectrum | Initial photon distribution | Comoving quasi-thermal/blackbody-like distribution |
| Injection spectrum | Fiducial photon temperature | $T_\gamma = 10^8$ K |
| Injection spectrum | Alternative option | Temperature estimated from local SRHD pressure and density when enabled |
| Photon statistics | Photon number per run | $2 \times 10^4$–$5 \times 10^5$ depending on the scan and test |
| Photon statistics | Fiducial production level | Main figures use the photon number of the corresponding scan; the highest-statistics runs reach $5 \times 10^5$ photons |
| Photon transport | Maximum number of scatterings | $n_{\rm scat,max} = 2000$ |
| Photon transport | Maximum ray-marching steps per leg | $10^5$ |
| Escape criterion | Escape threshold | $\tau_{\rm esc} = 1$ |
| Escape criterion | Acceptance cone | Photons recorded if they satisfy the observer-cone selection around $\theta_{\rm obs}$ |
| Observer setup | Viewing-angle scan | $\theta_{\rm obs} = 0°,\ 0.5°,\ 1°,\ 2°$ |
| Observer setup | Detector half-angle | $\theta_{\rm det} = 1°$ |
| Observer setup | Redshift | Fiducial synthetic calculations use $z = 0$ unless otherwise stated |
| Polarization | Stokes quantities stored | Packet weight $w$ and normalized Stokes parameters $(q, u)$ |
| Polarization estimator | Integrated polarization | $\Pi = \sqrt{Q^2 + U^2}/I$, $\chi = \frac{1}{2}\arctan(U/Q)$; $I = \sum w_i$, $Q = \sum w_i q_i$, $U = \sum w_i u_i$ |
| Adiabatic evolution | Between scatterings | Effective comoving scaling $E_\gamma' \propto r^{-2/3}$ |
| Electron sampling | Thermal electrons | Maxwell–Jüttner distribution in the comoving frame |
| Scattering physics | Interaction kernel | Klein–Nishina Compton scattering in the electron rest frame |
| Polarization update | Scattering treatment | Stokes vector update using the scattering-plane basis and Mueller matrix formalism |

**Table 3**
Summary of the Main Parameter Scans, Case Labels, and Derived Monte Carlo Observables

| Category | Quantity | Value/Description |
|---|---|---|
| Dissipation model | Reheating trigger window | $\tau_{\rm diss,min} \leqslant \tau \leqslant \tau_{\rm diss,max}$ |
| Dissipation model | Fiducial reference window | $(\tau_{\rm diss,min}, \tau_{\rm diss,max}) = (3, 30)$ |
| Dissipation scan | Window grid | $\tau_{\rm diss,min} = 1,\ 3,\ 10$ and $\tau_{\rm diss,max} = 20,\ 30,\ 50,\ 80$, retaining only $\tau_{\rm diss,max} > \tau_{\rm diss,min}$ |
| Dissipation model | Reheating cadence | Every $n_{\rm rh} = 5$ scatterings in the fiducial prescription |
| Dissipation model | Fiducial electron-temperature boost | $\Theta_{e,\rm boost} = 0.3$ |
| Pair-loading scan | Scanned values | $Z_\pm = 0,\ 0.3,\ 1,\ 2,\ 3$ |
| Baseline case labels | Dynamic reference cases | D64 and D74 |
| Baseline case labels | Axis reference cases | A68 and A74 |
| Case-label meaning | Interpretation | Case identifiers inherited from file naming; not literal injection-plane coordinates |
| Outputs | Photon-list columns | $t_{\rm obs}$, $E_{\rm obs}$, $w$, $q$, $u$, $\theta_{\rm obs}$, $r_{\rm last}$, $z_{\rm last}$, region tag, $n_{\rm scat}$ |
| Derived observables | Post-processed diagnostics | Light curves, $E_{\rm pk}(t)$, $\alpha_{\rm eff}(t)$, tail strength, $\Pi(t)$, $\chi(t)$, $\Pi(E)$, $\chi(E)$, and last-scattering statistics |

Analytical radiation-hydrodynamic models describe the coupled evolution of radiation and matter in an averaged sense through moment closure relations, typically valid under quasi-steady and near-spherical conditions. By contrast, the Monte Carlo approach explicitly tracks the scattering history of individual photons in a given SRHD background, thereby avoiding angular closure assumptions and naturally incorporating strong anisotropy, structured jets, and polarization effects (A. M. Beloborodov 2011; C. Lundman et al. 2014).

The purpose of this section is to detail the statistical implementation of photon propagation, Klein–Nishina Compton scattering, polarization (Stokes vector) updates, and the post-processing procedure for observers characterized by viewing angle $\theta_{\rm obs}$ and detector acceptance cone $\theta_{\rm det}$.

### *4.1. SRHD Background, Interpolation, and Electron Density*

*SRHD background and geometry.* Radiative transfer is performed on a two-dimensional axisymmetric SRHD

**Table 4**
Mapping from Code Units to Physical and Observer-frame Quantities Used in the SRHD + Monte Carlo Calculations

| Quantity | Definition/Value |
|---|---|
| Length unit | $L_0 = 10^{10}$ cm |
| Time unit | $t_0 = L_0/c \simeq 0.3336$ s |
| Density unit | $\rho_0 = 10^{-7}$ g cm$^{-3}$ |
| Pressure unit | Set consistently by the adopted code-unit normalization |
| Observer-frame energy output | Stored as $E_{\rm obs}$ in cgs units (erg) |
| Displayed spectral energy | Converted to keV or MeV in post-processing plots |
| Observer-frame arrival time | $t_{\rm obs} = (t - \hat{k} \cdot \mathbf{r}/c)/(1 + z)$ |
| Observer-frame photon energy | $E_{\rm obs} = E/(1 + z)$ |
| Peak-energy diagnostic | $E_{\rm pk}$ defined from the peak of $\nu F_\nu \propto E^2\, dN/dE$ |
| Low-energy slope proxy | $\alpha_{\rm eff}$ measured from a log–log fit below $E_{\rm pk}$ in post-processing |
| Tail-strength diagnostic | tail $\equiv \nu F_\nu(3E_{\rm pk})/\nu F_{\nu,{\rm pk}}$ |
| Polarization estimator | $\Pi = \sqrt{Q^2 + U^2}/I$, $\chi = \frac{1}{2}\arctan(U/Q)$ |

background. Each hydrodynamic snapshot provides cell-centered quantities $\rho(R_{\rm cyl}, z, t)$, $p(R_{\rm cyl}, z, t)$, and $\boldsymbol{v}(R_{\rm cyl}, z, t)$ (or equivalently $\Gamma$ and $\boldsymbol{\beta}$). Within each grid cell, the background is treated as piecewise constant. When a photon crosses a cell boundary, local quantities $\rho$, $p$, $\Gamma$, and $\boldsymbol{\beta}$ are updated accordingly (H. Ito et al. 2013; R. Santana et al. 2014).

For time-dependent simulations based on discrete snapshots $\{t_k\}$, the background at time $t \in [t_k, t_{k+1})$ is taken from snapshot $k$. Linear interpolation may alternatively be used to reduce temporal discretization error (W. H. Press et al. 1992); in this work we adopt the piecewise-constant scheme.

*Electron density and pair loading.* In SRHD, $\rho$ denotes the comoving rest-mass density. The comoving proton number density is therefore

$$n'_p(R_{\rm cyl}, z, t) = \frac{\rho(R_{\rm cyl}, z, t)}{m_p}. \tag{13}$$

If pair production is included, the total comoving electron number density becomes

$$n'_e(R_{\rm cyl}, z, t) = (1 + Z_\pm)\frac{\rho(R_{\rm cyl}, z, t)}{m_p}, \tag{14}$$

where $Z_\pm$ may be treated as a constant parameter or supplied by a semianalytical model. In this work, $Z_\pm$ is treated as a controllable parameter to investigate its impact on the photospheric radius, spectral shape, and polarization.

### 4.2. Injection Conditions: Injection Surface, Initial Photons, and Stokes Parameters

*Injection surface.* Photons are injected in regions of high optical depth where the radiation field is approximately thermal (A. Pe'er 2008; A. M. Beloborodov 2011). In the SRHD background, the injection surface is defined geometrically (e.g., a plane at $z = z_{\rm init}$ within $R_{\rm cyl} \leqslant R_{\rm inj}$). The outward optical depth $\tau_{\rm out}$ is computed dynamically to ensure $\tau_{\rm out} \gg 1$, guaranteeing injection in an optically thick region.

*Initial energy, direction, and Stokes vector.* Photon energies are sampled in the local comoving frame from a blackbody (or quasi-thermal) distribution (G. B. Rybicki & A. P. Lightman 1979):

$$p(E'_\gamma) \propto \frac{E'^2_\gamma}{\exp(E'_\gamma/kT'_\gamma) - 1}. \tag{15}$$

Initial directions are isotropic in the comoving frame. Photons are initialized as unpolarized with Stokes vector

$$\mathbf{S} = (I, Q, U)^{\rm T}, \qquad Q = U = 0, \tag{16}$$

neglecting circular polarization. Each photon packet carries a statistical weight $w$.

### 4.3. Optical-depth Sampling and Dynamic Determination of Interaction Sites

For each scattering event, the cumulative optical depth is sampled from

$$\Delta\tau = -\ln\xi, \qquad \xi \in (0, 1), \tag{17}$$

corresponding to the exponential free-path distribution in optical-depth space (L. A. Pozdnyakov et al. 1983).

Photon propagation proceeds via ray marching in the laboratory frame. Within each grid cell, local quantities are treated as constant and the optical-depth increment is

$$d\tau = n'_e\, \sigma_{\rm T}\, ds', \qquad ds' = \Gamma(1 - \boldsymbol{\beta} \cdot \hat{\Omega})\, ds, \tag{18}$$

where $ds$ is the lab-frame path length and $ds'$ is the corresponding comoving interval (G. B. Rybicki & A. P. Lightman 1979).

### 4.4. Escape Criterion

After each scattering, the remaining optical depth along the outgoing direction is computed as

$$\tau_{\rm out}(\hat{\Omega}_{\rm out}) = \int_{\rm ray} n'_e\, \sigma_{\rm T}\, ds'. \tag{19}$$

If $\tau_{\rm out} < \tau_{\rm esc}$ (with $\tau_{\rm esc} \sim 1$), the photon is considered decoupled and recorded (A. Pe'er 2008; A. M. Beloborodov 2011; H. Ito et al. 2013; C. Lundman et al. 2014); otherwise, the next scattering cycle is initiated.

### 4.5. Adiabatic Cooling between Scatterings

Between successive scatterings, photons propagate through an expanding SRHD background. In addition to the discrete energy exchange at Compton-scattering events, the comoving photon energy is expected to decrease as the background flow expands. In the present work, this effect is incorporated through an approximate post-processing treatment of adiabatic cooling.

Specifically, we do not evolve the photon field with a fully self-consistent radiation-hydrodynamic energy equation between scatterings. Instead, we apply an effective prescription intended to capture the leading influence of expansion in the prescribed SRHD outflow on the comoving photon energy. The adiabatic term should therefore be interpreted as an approximate description tied to the expanding hydrodynamic background, rather than as a self-consistent dynamical coupling between radiation and fluid.

For a relativistic expanding outflow, we adopt the effective scaling

$$E'_\gamma \propto r^{-2/3}, \tag{20}$$

which is implemented numerically as

$$E'_{\gamma,\mathrm{new}} = E'_{\gamma,\mathrm{old}}\left(\frac{r_{\mathrm{new}}}{r_{\mathrm{old}}}\right)^{-2/3}. \tag{21}$$

Here $E'_\gamma$ denotes the photon energy in the comoving frame, and $r_{\mathrm{old}}$ and $r_{\mathrm{new}}$ are the radii of the packet before and after free propagation between two successive scattering events. This prescription provides a simple representation of expansion-induced softening while keeping the Monte Carlo transport numerically tractable over many scatterings and across multiple SRHD snapshots.

We emphasize that this treatment is approximate. The hydrodynamic background is evolved independently of the radiation field, and radiative feedback on the flow is neglected. Accordingly, the present method should be regarded as a radiative-transfer post-processing calculation on a prescribed, time-dependent SRHD background. Its role is to capture the dominant expansion-related energy loss between scatterings in a physically motivated but simplified manner, following common practice in photospheric post-processing treatments (A. M. Beloborodov 2011; H. Ito et al. 2013). The principal conclusions of this paper are based on relative trends across the parameter scans and dynamic–axis comparisons and are therefore expected to be more robust than the absolute normalization of any single spectral quantity.

### 4.6. Electron Sampling and Scattering in the Electron Rest Frame

At each interaction, electron velocities are sampled from a Maxwell–Jüttner distribution in the comoving frame (F. Jüttner 1911):

$$f(\gamma_e)\,d\gamma_e \propto \gamma_e^2\beta_e \exp\left(-\frac{\gamma_e}{\Theta_e}\right)d\gamma_e, \qquad \Theta_e = \frac{kT'_e}{m_e c^2}. \tag{22}$$

The scattering is performed in the electron rest frame and transformed back to the comoving and laboratory frames.

### 4.7. Klein–Nishina Sampling and Polarization Update

In the electron rest frame, the scattered photon energy satisfies

$$\varepsilon_f = \frac{\varepsilon_i}{1 + \varepsilon_i(1 - \cos\theta_{\mathrm{sc}})}, \tag{23}$$

where $\varepsilon_i = E_i/(m_e c^2)$. Scattering angles are sampled from the Klein–Nishina differential cross section using rejection sampling (A. H. Compton 1923; O. Klein & Y. Nishina 1929; L. A. Pozdnyakov et al. 1983).

Polarization evolution is treated using Stokes vectors. After rotating into the scattering-plane basis, the Stokes vector is updated via the Mueller matrix:

$$\boldsymbol{S}_{\mathrm{out}} = \boldsymbol{R}(-\psi)\,\boldsymbol{M}_{\mathrm{KN}}(\varepsilon_i, \theta_{\mathrm{sc}})\,\boldsymbol{R}(\psi)\,\boldsymbol{S}_{\mathrm{in}}. \tag{24}$$

### 4.8. Subphotospheric Reheating

To model subphotospheric energy dissipation and unsaturated Comptonization, we introduce parameterized reheating (A. M. Beloborodov 2010; I. Vurm & A. M. Beloborodov 2016). If a scattering event occurs within

$$\tau_{\mathrm{diss,min}} \leqslant \tau_{\mathrm{out}} \leqslant \tau_{\mathrm{diss,max}} \tag{25}$$

and the scattering count satisfies $n_{\mathrm{scat}} \bmod n_{\mathrm{rh}} = 0$, the electron temperature is incremented to maintain $y \sim \mathcal{O}(1)$.

### 4.9. Observer Post-processing

*Arrival time and energy.* The observed arrival time is computed using equal-arrival-time geometry

$$t_{\mathrm{obs}} = \frac{1}{1+z}\left(t - \frac{\hat{k}\cdot\mathbf{r}}{c}\right) \tag{26}$$

(J. Granot et al. 1999), with observed energy $E_{\mathrm{obs}} = E/(1+z)$.

*Time-resolved spectra.* Within each time bin, spectra are constructed and fitted with the Band function (D. Band et al. 1993) to obtain ($\alpha$, $\beta$, $E_{\mathrm{pk}}$).

*Polarization.* Stokes parameters are summed within bins:

$$I_{\mathrm{bin}} = \sum_i w_i, \qquad Q_{\mathrm{bin}} = \sum_i w_i q_i, \qquad U_{\mathrm{bin}} = \sum_i w_i u_i, \tag{27}$$

yielding polarization degree and angle,

$$\Pi = \frac{\sqrt{Q_{\mathrm{bin}}^2 + U_{\mathrm{bin}}^2}}{I_{\mathrm{bin}}}, \qquad \chi = \frac{1}{2}\arctan\left(\frac{U_{\mathrm{bin}}}{Q_{\mathrm{bin}}}\right). \tag{28}$$

This definition follows the standard post-processing procedure for scattering-induced polarization (S. Chandrasekhar 1960; C. Lundman et al. 2014).

### 4.10. Run Configuration and Parameter Ranges

For reproducibility, the fiducial SRHD setup is summarized in Table 1, the Monte Carlo transport setup and parameter scans are listed in Tables 2 and 3, and the mapping between code units and observer-frame quantities is given in Table 4. Unless otherwise stated, the SRHD background is evolved in two-dimensional axisymmetric cylindrical coordinates, and the Monte Carlo radiative transfer is performed on the resulting time-ordered snapshot sequence using a piecewise-constant-in-time approximation.

**Table 5**
Physical Setup of the Representative Baseline Calculations Shown in Figure 5

| Case | Treatment | $z_{\rm init}$ | $z_{\rm init}$ (cm) | $\tau_{\rm init}$ (0) | $\tau_{\rm init}$ ($0.5R_{\rm inj}$) | $\tau_{\rm init}$ ($R_{\rm inj}$) | Injected/Kept |
|---|---|---|---|---|---|---|---|
| A74 | axis reference | 74 | $7.4 \times 10^{11}$ | 32.45 | 31.12 | 25.66 | 20,000/11,268 |
| D74 | dynamic | 74 | $7.4 \times 10^{11}$ | 32.45 | 31.12 | 25.66 | 20,000/14,474 |
| D64 | dynamic | 64 | $6.4 \times 10^{11}$ | 110.27 | 104.97 | 84.54 | 20,000/14,592 |

**Note.** The quoted optical depths are evaluated on the first SRHD frame used in the Monte Carlo sequence, along the on-axis observer direction, at the symmetry axis, at the diagnostic radius $0.5R_{\rm inj}$ used in the code, and at the edge of the injection disk.

The calculations presented below include two optical-depth treatments. In the dynamic runs, photon escape is determined from the full direction-dependent residual optical depth, $\tau_{\rm out}(\hat{\Omega})$, evaluated along the actual photon-propagation direction. In parallel, we also perform axis-based reference runs using a simplified optical-depth prescription. The dynamic calculations are regarded as the main results of this work, whereas the axis runs serve as reduced comparison models that help isolate which radiative trends are robust and which depend on the full time-dependent structured decoupling geometry. For the axis-based reference runs, this reduced prescription is implemented through a one-dimensional on-axis profile $\tau_{\rm axis}(z)$, computed from the first SRHD snapshot by integrating the Thomson optical depth from $(R_{\rm cyl}, z) = (0, z)$ in the $+z$-direction. The tabulated profile is used as the escape criterion, $\tau_{\rm axis}(z) < \tau_{\rm esc} = 1$, while photon propagation, free-path sampling, scattering, energy exchange, and Stokes vector updates are still performed in the SRHD background. Thus, the axis runs isolate the effect of replacing the full path-dependent decoupling criterion by an on-axis optical-depth reference.

The principal parameter scans considered in this section are as follows. The dissipation-window scan uses $\tau_{\rm diss,min} = 1,\ 3,\ 10$ and $\tau_{\rm diss,max} = 20,\ 30,\ 50,\ 80$, retaining only combinations with $\tau_{\rm diss,max} > \tau_{\rm diss,min}$. The pair-loading scan adopts $Z_\pm = 0,\ 0.3,\ 1,\ 2,$ and 3. The fiducial viewing-angle sequence is $\theta_{\rm obs} = 0°,\ 0.5°,\ 1°,$ and $2°$. Photon statistics depend on the specific scan and test, spanning approximately $2 \times 10^4$ to $5 \times 10^5$ photon packets; unless otherwise stated, each displayed time bin contains at least $\sim$200 escaping photons.

Case labels follow the original output-file convention: the leading letter denotes the optical-depth treatment, with D for the full dynamic calculation and A for the axis-based reference calculation, while the number gives the injection-plane coordinate $z_{\rm init}$ in code units. The physical setup and initial optical depths of the representative baseline cases are listed in Table 5.

Unless otherwise stated, the synthetic calculations shown in this section adopt $z = 0$ so that the reported trends reflect the internal transfer physics directly. Cosmological redshift can be incorporated straightforwardly in post-processing through the standard transformations listed in Table 4.

## 5. Results

### 5.1. Comparison Strategy and Scope

In this section we present the Monte Carlo radiative-transfer results obtained on top of the time-dependent SRHD jet backgrounds described in Sections 3 and 4. The corresponding numerical configurations and parameter ranges have been summarized earlier in Section 4.10.

The main purpose of this section is not only to present the radiative-transfer outputs themselves but also to compare two different optical-depth treatments in a systematic way. The first is the *dynamic* treatment, in which photon escape and related diagnostics are determined from the full direction-dependent residual optical depth, $\tau_{\rm out}(\hat{\Omega})$, evaluated along the actual photon-propagation direction in the time-dependent structured SRHD background. The second is an *axis-based* reference treatment, which serves as a reduced comparison model with a simplified optical-depth prescription.

Throughout this section, the dynamic calculations are regarded as the main results of this work, while the axis runs are used as a reference baseline to identify which radiative trends are robust and which are more sensitive to the full path-dependent decoupling geometry. Accordingly, the discussion below is organized around three main comparisons: the baseline no-dissipation case, the dissipation-window scan, and the pair-loading scan. This comparison strategy is then summarized in terms of geometry-sensitive versus geometry-robust observables and interpreted in light of the last-scattering statistics.

### 5.2. Baseline Comparison

We begin with a baseline model in which no subphotospheric dissipation is included and the pair-loading parameter is set to $Z_\pm = 0$. Photons are injected at large optical depths with a comoving quasi-thermal distribution and subsequently decouple gradually through multiple Compton scatterings off thermal electrons, together with geometric expansion. No additional energy injection or nonthermal particle acceleration is introduced.

In this baseline comparison, we contrast two realizations of the same no-dissipation, no-pair setup: (i) the full time-dependent Monte Carlo radiative transfer on the structured SRHD background with the direction-dependent escape criterion $\tau_{\rm out}(\hat{\Omega})$, and (ii) the corresponding axis-based reference calculation, which provides a simplified optical-depth baseline. The goal is to determine which features of the quasi-thermal photospheric baseline are already recovered in the reduced treatment and which depend more sensitively on the full time-dependent structured decoupling geometry.

The baseline cases shown below use the same no-dissipation, no-pair physical assumptions but different injection planes and optical-depth treatments. Photons are injected uniformly over a disk of radius $R_{\rm inj} = 2.0$ code units with a comoving quasi-thermal spectrum of temperature $T_\gamma = 10^8$ K and with $Z_\pm = 0$. With $L_0 = 10^{10}$ cm, D64 and D74 correspond to $z_{\rm init} = 6.4 \times 10^{11}$ and $7.4 \times 10^{11}$ cm, respectively. In the D74/A74 setup the initial optical depth across the injection disk is $\tau_{\rm init} \simeq$ 26–32, while in the D64 setup it is $\tau_{\rm init} \simeq$ 85–110.

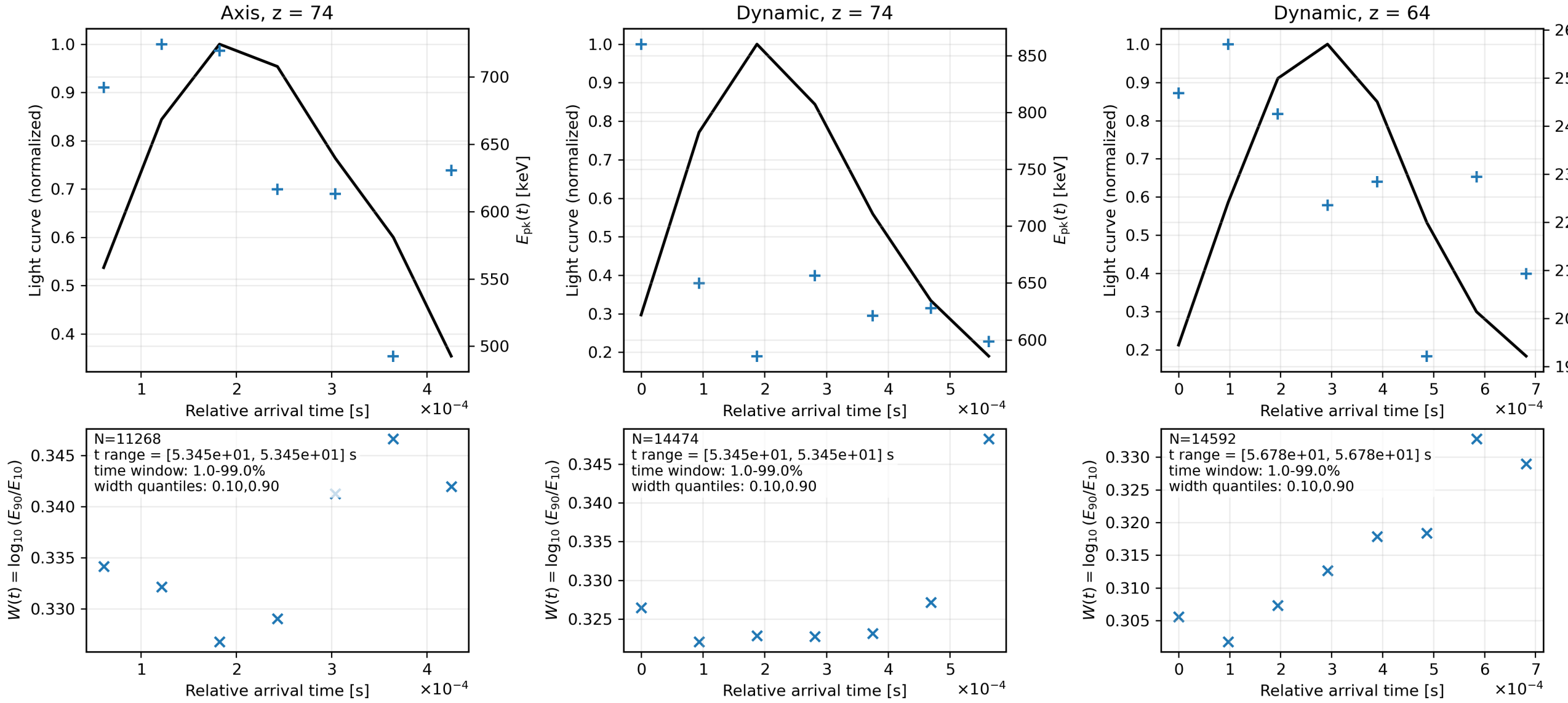

**Figure 5.** Baseline comparison among three representative no-dissipation, no-pair cases. From left to right, the columns show the axis-based reference calculation A74, the dynamic calculation D74, and the dynamic calculation D64. The top panels show the normalized light curve (black line) and the time-resolved peak energy $E_{\rm pk}(t)$ (symbols; right axis, in keV). The bottom panels show the spectral width $W(t) = \log_{10}(E_{90}/E_{10})$. The comparison between the first and second columns isolates the effect of the optical-depth treatment at fixed injection depth, while the comparison between the second and third columns shows the weak sensitivity of the dynamic baseline results to injection depth. The corresponding injection optical depths are listed in Table 5. The displayed time axis corresponds to the main photon-arrival window used for time-resolved analysis, rather than the full formal arrival-time extent of all escaped photons.

Figure 5 summarizes three representative baseline realizations in a single layout: the axis-based reference calculation A74, the dynamic calculation D74, and the dynamic calculation D64. The top panels show the normalized light curve (black line), together with the time-resolved peak energy $E_{\rm pk}(t)$ (symbols; right axis, in keV). The bottom panels show the spectral width

$$W(t) = \log_{10}(E_{90}/E_{10}), \tag{29}$$

where $E_{10}$ and $E_{90}$ enclose 10% and 90% of the cumulative $\nu F_\nu$ distribution, respectively.

For clarity, the time-resolved quantities in Figure 5 are evaluated over the dominant photon-arrival window rather than over the full formal arrival-time extent of all escaped photons. This choice is motivated by photon statistics: the sparsely populated arrival-time tails, especially in the dynamic treatment, are not suitable for stable measurements of $E_{\rm pk}(t)$ and $W(t)$. Accordingly, the horizontal axis in this figure should be interpreted as the relative arrival time within the main emission window used for time-resolved analysis, rather than as the total burst duration of the full synthetic signal.

The comparison between the first and second columns isolates the effect of the optical-depth treatment at fixed injection depth. At $z_{\rm init} = 74$ (A74 and D74), both the axis and dynamic calculations produce a smooth single-peaked pulse, only moderate temporal evolution of $E_{\rm pk}(t)$, and a narrow spectral width. This shows that the leading-order baseline behavior is robust between the two treatments. In particular, the quasi-thermal character of the baseline spectrum is not an artifact of the full dynamic treatment alone.

The comparison between the second and third columns isolates the sensitivity to injection depth within the dynamic treatment. Despite the different initial optical-depth conditions associated with D74 and D64, the dynamic light curves remain broadly similar in morphology, the temporal variation of $E_{\rm pk}(t)$ remains moderate, and the spectral width stays narrow, with typical values of $W \simeq$ 0.30–0.35. This indicates that, at baseline level, the radiation properties depend only weakly on the injection depth and therefore provide a stable reference for the more diagnostic comparisons that follow.

More generally, the baseline model shows that, in the absence of subphotospheric dissipation and pair production, multiple scatterings and geometric expansion smooth the spectrum but do not drive it far from a thermal-like shape. The resulting pulse is simple, the peak-energy evolution is limited, and the spectral width remains substantially narrower than that typically inferred for observed GRB Band spectra. In this sense, the baseline photospheric behavior is one of the least geometry-sensitive parts of the problem.

At the same time, the agreement between the axis and dynamic treatments is not exact. In the dynamic calculation, photon decoupling occurs in a genuinely structured and time-dependent background, so the time-resolved quantities still reflect the broader decoupling geometry encoded by $\tau_{\rm out}(\hat{\Omega})$. By contrast, the axis-based result provides a reduced reference in which the same leading quasi-thermal baseline is preserved but the full spatiotemporal complexity of the structured flow is absent. This distinction is modest at baseline level, but it becomes important when we later examine dissipation-depth effects, pair loading, and the geometry sensitivity of polarization.

### *5.3. Dissipation-window Comparison*

To investigate how subphotospheric energy dissipation shapes the photospheric spectrum, we introduce a parameterized reheating prescription that operates only within a finite optical-depth window, $\tau_{\rm diss,min} \leqslant \tau \leqslant \tau_{\rm diss,max}$. Within this window, the electron temperature is periodically boosted in a subset of scattering events, maintaining a nonzero Compton $y$-parameter. This prescription is deliberately parameterized. We do not compute the microscopic dissipation rate from shocks, reconnection, or radiative feedback. Instead, within the selected optical-depth window we impose a constant per-event reheating strength, implemented in the fiducial dissipative runs as a fixed electron-temperature boost $\Delta\Theta_e = 0.3$ every $n_{rh} = 5$ scatterings. Outside this window no additional heating is applied. The following scan therefore varies the location and width of the dissipation region in optical-depth space, while holding the local reheating prescription fixed. In the baseline calculations discussed above, the reheating channel is switched off.

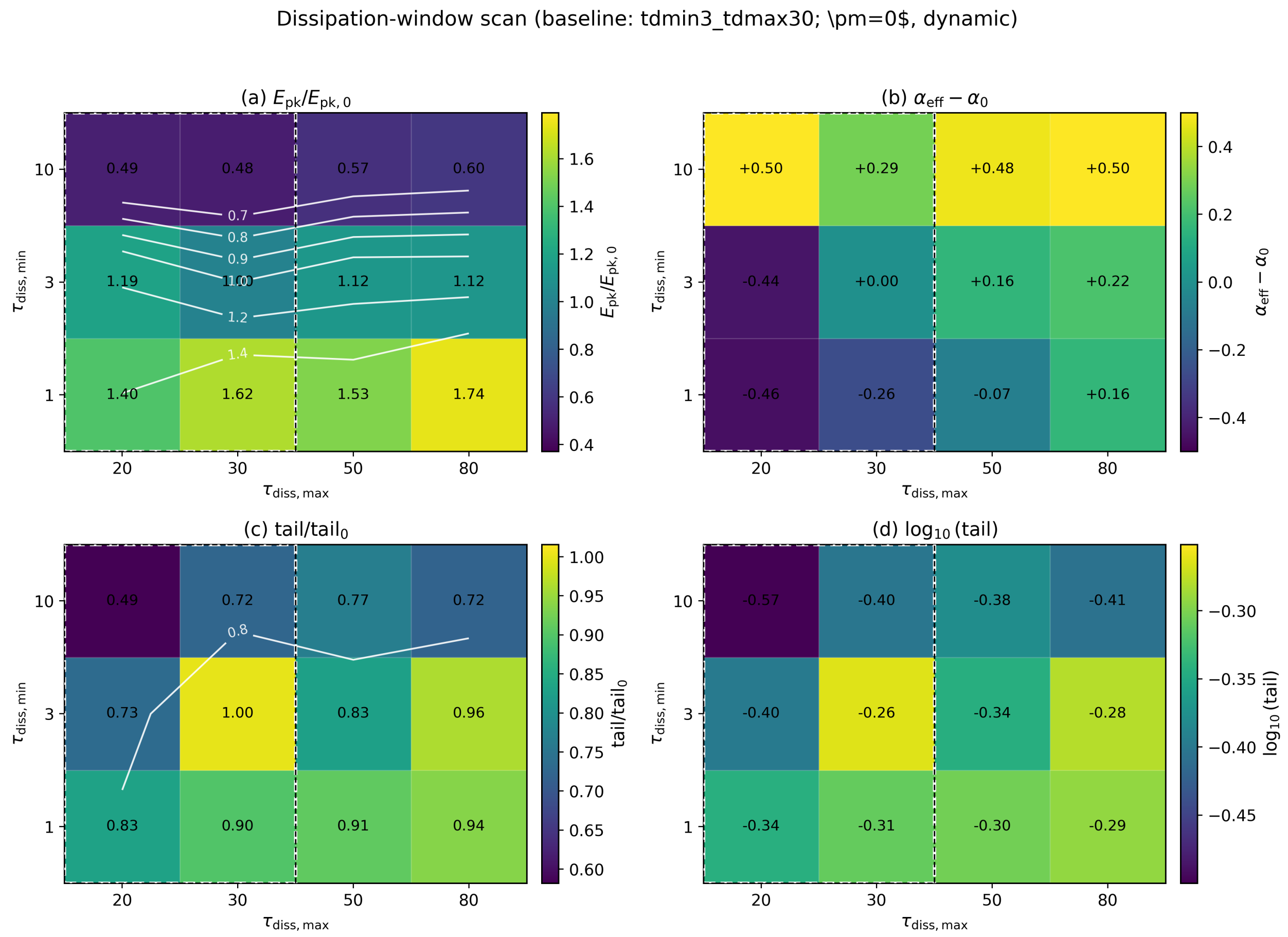

**Figure 6.** Dynamic two-dimensional phase space for the optical-depth window scan of subphotospheric dissipation. From left to right and top to bottom, the panels show the relative change in peak energy $E_{pk}/E_{pk,0}$, the change in low-energy spectral index $\alpha_{eff} - \alpha_0$, the relative high-energy tail strength tail/tail$_0$, and log$_{10}$(tail), as functions of ($\tau_{diss,min}$, $\tau_{diss,max}$). All quantities are normalized to the reference dissipation model ($\tau_{diss,min}$, $\tau_{diss,max}$) = (3, 30), marked by the dashed lines.

In this subsection we compare two implementations of the dissipation-window grid summarized in Section 4.10. The first is the full dynamic calculation, in which both escape and dissipation are evaluated using the direction-dependent residual optical depth along the actual photon-propagation direction, $\tau_{out}(\hat{\Omega})$. The second is the axis-based reference treatment described in Section 4.10.

For both treatments, the phase-space diagnostics are constructed from the same post-processing definitions of $E_{pk}$, $\alpha_{eff}$, and tail strength, and each panel is normalized to the same formal reference point ($\tau_{diss,min}$, $\tau_{diss,max}$) = (3, 30) within the corresponding treatment. The dynamic results are shown in Figure 6, while the corresponding axis-based reference results are shown in Figure 7.

The dynamic results show that the spectral response is highly sensitive to where dissipation occurs in optical depth and that the parameter space separates into distinct regimes. When dissipation occurs at shallower depths, especially for $\tau_{diss,min} \simeq 1$, the peak energy increases relative to the reference case. In this regime, energy injection occurs before substantial adiabatic cooling, raising the mean photon energy. At the same time, $\alpha_{eff}$ decreases and the tail-related diagnostics strengthen, indicating that nonsaturated Comptonization becomes increasingly important both below and above the spectral peak.

By contrast, when dissipation is confined to deeper regions, particularly for $\tau_{diss,min} \gtrsim 10$, the peak energy is systematically reduced relative to the shallow-dissipation cases. In this case, photons experience more post-heating scatterings before escape, leading to stronger thermalization and stronger adiabatic cooling. The high-energy tail becomes correspondingly weaker, and the spectrum approaches a more quasi-thermal shape.

Between these two extremes, an intermediate optical-depth window, typically around $\tau_{diss,min} \sim 3$ and $\tau_{diss,max} \sim 20$–50, yields the most efficient nonsaturated Comptonization. In this regime, $E_{pk}$ remains closer to the reference value, while the high-energy tail is enhanced and $\alpha_{eff}$ changes more moderately, implying substantial spectral broadening without full thermalization. Taken together, the dynamic scan shows that the photospheric spectral shape is controlled not only by the presence of dissipation but also crucially by the optical-depth range over which dissipation operates.

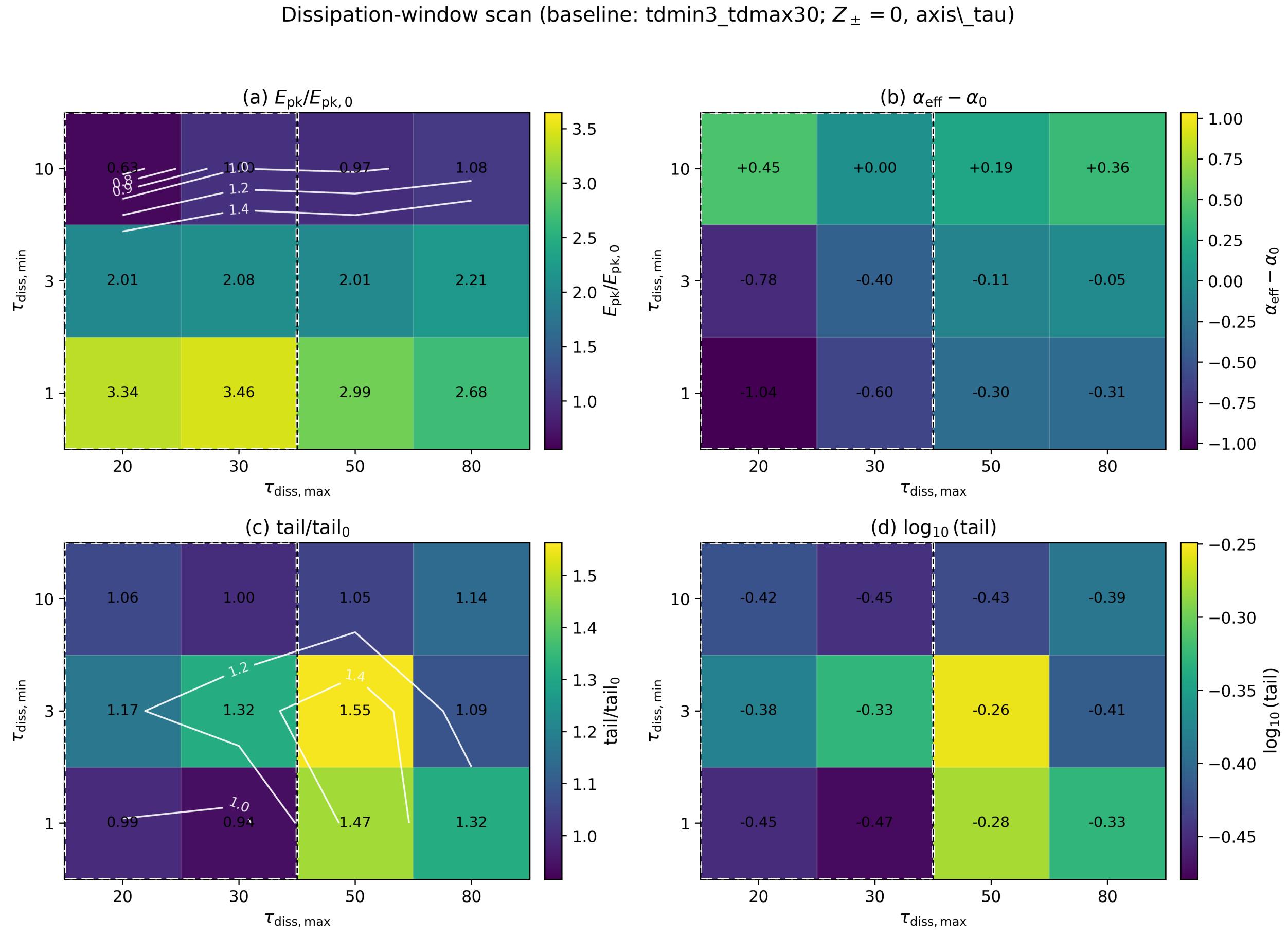

**Figure 7.** Axis-based reference two-dimensional phase space for the same optical-depth window scan shown in Figure 6. The same diagnostics, panel ordering, and normalization conventions are used. Comparison with the dynamic treatment shows that the leading ordering of the dissipation-depth response is qualitatively robust, while the quantitative amplitudes differ because the effective reheating geometry is simplified in the axis treatment.

The comparison with the axis-based reference calculation shows that this overall ordering of the dissipation-window parameter space is robust. In both treatments, shallow dissipation tends to harden the spectrum and increase $E_{pk}$, deep dissipation tends to drive the radiation toward stronger thermalization, and an intermediate optical-depth window is most favorable for producing substantial spectral broadening. This agreement indicates that the first-order dependence of the spectrum on dissipation depth is not an artifact of one specific optical-depth prescription.

At the same time, the quantitative responses are not identical. The dynamic calculation generally shows a different amplitude of variation across the phase space than the axis reference. This is physically expected because the same formal dissipation window does not correspond to the same effective reheating geometry in the two models. In the dynamic treatment, dissipation is evaluated using the actual direction-dependent residual optical depth, so the reheated photons sample the full structured SRHD background, together with its angular gradients, shear layers, and time-dependent last-scattering geometry.

This difference is especially important when interpreting the tail metric. In our post-processing, the tail strength is measured relative to the spectral peak, so it depends not only on the formation of the nonthermal tail itself but also on the shift of $E_{pk}$ across the parameter space. Differences between the dynamic and axis panels in the tail-related quantities should therefore be interpreted as a combined consequence of tail formation and peak-energy evolution, rather than as a pure change in the far high-energy slope alone.

Overall, the dissipation-window comparison leads to two conclusions. First, the qualitative spectral ordering of the parameter space is robust: the observed $E_{pk}$, low-energy slope, and relative tail strength all retain a clear sensitivity to the depth at which dissipation operates. Second, the full dynamic treatment is required for quantitatively reliable predictions in a structured, time-dependent jet, because the effective reheating geometry depends on the full path-dependent decoupling history. We therefore interpret the axis scan as a useful reduced reference model and the dynamic scan as the more complete prediction of how dissipation depth is imprinted on observable spectra.

### 5.4. *Pair-loading Comparison*

We vary the electron–positron pair-loading parameter $Z_\pm$ while keeping all other conditions fixed, and we quantify its impact on the time-integrated spectral properties and polarization. In this subsection, we directly compare the full dynamic calculation with the corresponding axis-based reference treatment using the same post-processing definitions for all diagnostics. The scanned values are $Z_\pm = 0,\ 0.3,\ 1,\ 2,$ and $3$, and all quantities are normalized to the no-pair case within each treatment.

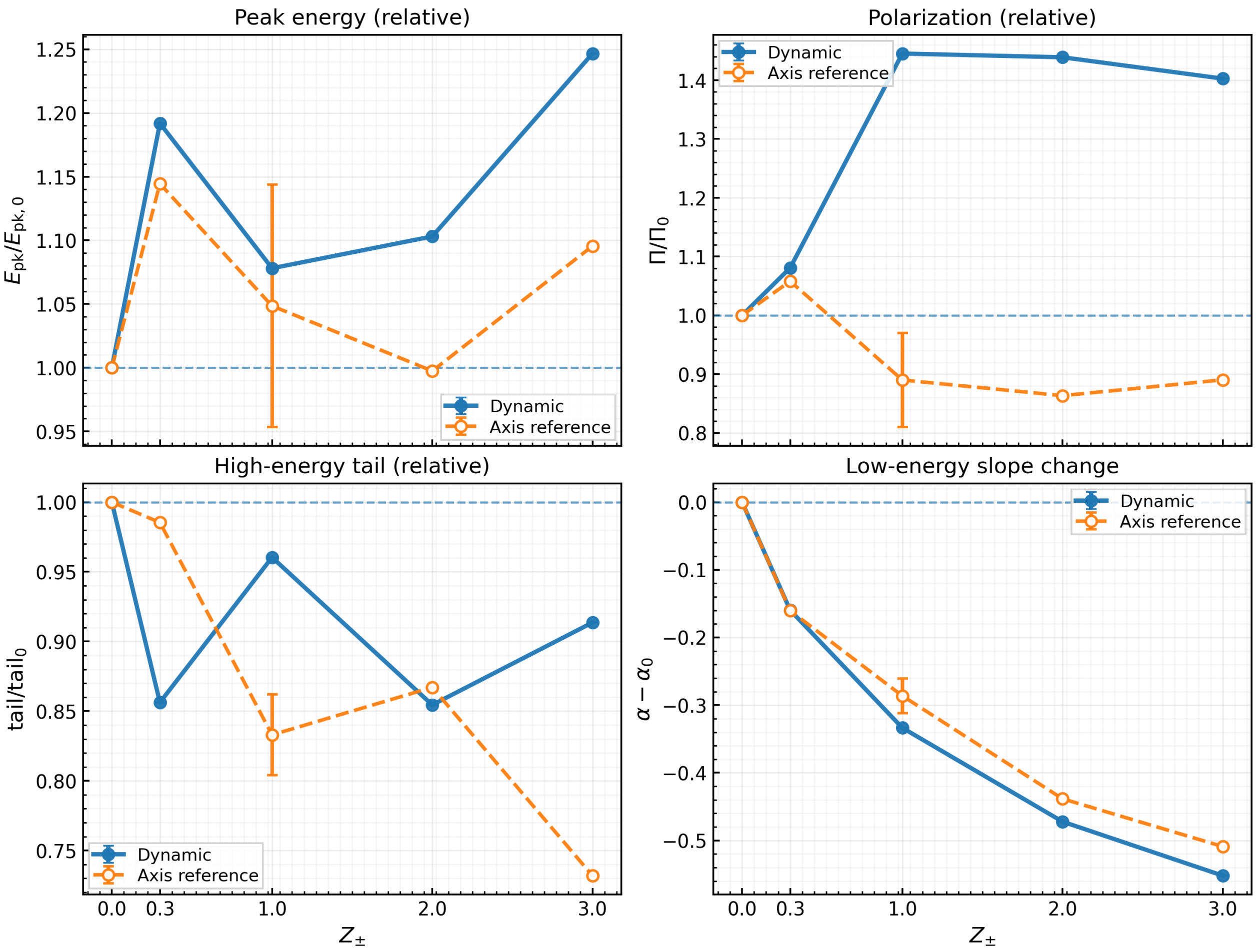

**Figure 8.** Comparison of the pair-loading response between the dynamic and axis-based treatments. The four panels show the relative evolution of $E_{pk}/E_{pk,0}$, $\Pi/\Pi_0$, the relative high-energy tail strength, and $\alpha_{eff} - \alpha_0$ as functions of $Z_\pm$. In all panels, the dynamic and axis results are normalized to the no-pair case within the corresponding treatment. The comparison shows that the main spectral trends with $Z_\pm$ are qualitatively robust between the two treatments, while the polarization response differs much more strongly, indicating that polarization is especially sensitive to the detailed geometry of photon decoupling.

Figure 8 summarizes the pair-loading response in the two treatments. The four panels show the relative evolution of the peak energy $E_{pk}/E_{pk,0}$, the polarization degree $\Pi/\Pi_0$, the relative high-energy tail strength, and the change in the low-energy spectral index $\alpha_{eff} - \alpha_0$, as functions of $Z_\pm$. Displaying the dynamic and axis results in the same panels makes it possible to identify immediately which trends are shared by both treatments and which depend more sensitively on the full path-dependent decoupling geometry.

First, the peak energy $E_{pk}$ increases overall with $Z_\pm$ in both calculations. Although a mild nonmonotonic behavior appears at intermediate pair loading, the highest-$Z_\pm$ cases remain above the no-pair baseline. In the dynamic calculation, the high pair-loading case reaches $E_{pk} \simeq 1.25\,E_{pk,0}$, whereas the axis-based reference shows a more modest enhancement, remaining closer to $\sim 1.10\,E_{pk,0}$. This common upward tendency indicates that pair loading generally strengthens the coupling between radiation and the plasma and prolongs energy exchange, thereby delaying decoupling and increasing the characteristic observed peak energy. The larger amplitude in the dynamic calculation suggests that, once the full direction-dependent decoupling geometry is retained, pair-supported scattering layers have a stronger effect on the photon histories that dominate the escaping signal.

The high-energy tail strength decreases overall as $Z_\pm$ increases, but the trend is not strictly monotonic. Compared to $Z_\pm = 0$, the tail is already suppressed at $Z_\pm \approx 0.3$, partially recovers near intermediate $Z_\pm$, and then weakens again toward the highest pair-loading end. This behavior is seen in both the dynamic and axis calculations, indicating that the suppression of the relative high-energy spectral strength is a robust consequence of stronger radiative coupling and increased effective scattering depth. At the same time, the nonmonotonic recovery at intermediate $Z_\pm$ implies that high-energy photon production and escape remain sensitive to the detailed competition among reheating, multiple scattering, and geometric decoupling.

The low-energy spectral index shows the clearest systematic evolution. In both treatments, $\alpha_{eff}$ decreases monotonically with $Z_\pm$, reaching $\alpha - \alpha_0 \simeq -0.55$ at $Z_\pm = 3$ in the dynamic case and a very similar value in the axis reference. This corresponds to a progressively steeper subpeak spectrum and can be attributed to the increased effective number of scatterings in the presence of additional pairs, which drives the low-energy photons closer to thermal equilibrium in the comoving frame. The close agreement between the dynamic and axis calculations for this trend indicates that the monotonic softening of the low-energy spectral slope is one of the most robust signatures of increasing pair loading in our model setup.

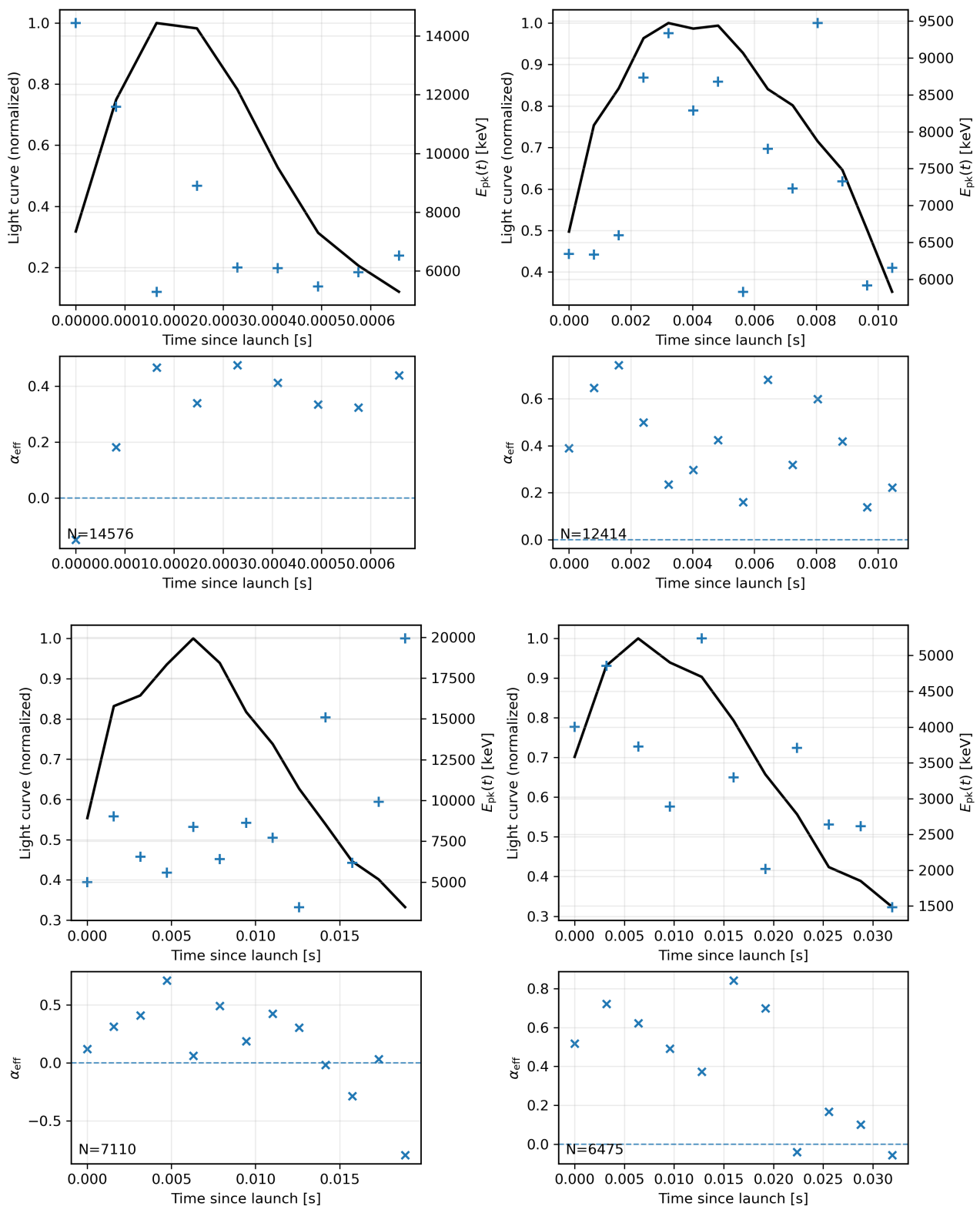

**Figure 9.** Viewing-angle dependence of time-resolved radiation properties. Top panel: on-axis and mildly off-axis views (0°, 0.5°). Bottom panel: off-axis views (1°, 2°). The panels show the normalized light curves, the time-resolved peak energy $E_{pk}(t)$, and the effective low-energy spectral index $\alpha_{eff}(t)$ for different viewing angles. The horizontal axis denotes the relative arrival time within the selected main photon-arrival window used for time-resolved analysis, rather than the full formal arrival-time extent of all escaped photons.

Polarization shows the largest difference between the two treatments. In the dynamic calculation, the time-integrated polarization degree increases markedly with $Z_{\pm}$, peaking around $Z_{\pm} \sim 1$–2 with $\Pi/\Pi_0 \simeq 1.4$–1.45, and then decreases slightly at higher $Z_{\pm}$ while remaining above the no-pair level. By contrast, in the axis-based reference calculation, $\Pi/\Pi_0$ remains close to unity at low $Z_{\pm}$ and then decreases below the no-pair value at larger pair loading. This contrast is physically informative. It shows that the polarization response to pair loading is much more sensitive than the spectral response to how the decoupling geometry is modeled. In the full dynamic treatment, additional pairs can reshape the angular distribution and anisotropy of the last-scattering region in a way that enhances the net polarization over a substantial range of $Z_{\pm}$. In the axis-based treatment, where the optical-depth geometry is reduced to a simplified reference description, this anisotropy enhancement is much weaker, and the dominant effect of increasing pair loading is instead to isotropize the radiation field and suppress the net polarization.

The pair-loading scan therefore leads to a mixed but physically useful conclusion. For spectral quantities such as $E_{pk}$, the high-energy tail, and especially $\alpha_{eff}$, the dynamic and axis treatments show the same leading ordering with $Z_{\pm}$, differing mainly in quantitative amplitude. For polarization, however, the two treatments diverge qualitatively. We interpret this as evidence that polarization is a more geometry-sensitive diagnostic of pair loading than the time-integrated spectral metrics. Equivalently, axis-based calculations are adequate for identifying the first-order spectral response to $Z_{\pm}$, but the full dynamic transfer treatment is required if one aims to predict how pair loading modifies the anisotropy of the last-scattering surface and hence the observable polarization.

In summary, pair loading modulates the effective scattering depth and decoupling history, systematically affecting the peak energy, the subpeak spectral slope, and the high-energy tail, while also producing a pronounced impact on polarization. The dynamic–axis comparison further clarifies that the spectral trends with $Z_{\pm}$ are relatively robust, whereas the polarization trend is strongly model dependent and carries direct information about the geometric realism of the transfer calculation. This makes polarization particularly valuable for discriminating between simplified and fully direction-dependent treatments of subphotospheric radiative transfer in structured jets.

### 5.5. Viewing-angle Dependence

Because the SRHD jet exhibits strong angular structure and velocity shear, the observed radiation properties depend systematically on the viewing angle $\theta_{obs}$. This dependence mainly arises from two effects: (i) the reduction of the Doppler factor with increasing $\theta_{obs}$, which causes deboosting and arrival-time delays, and (ii) the increasing geometric asymmetry of the last-scattering surface off-axis, which enhances radiation anisotropy and alters spectral evolution. We therefore perform synthetic observations from on-axis to off-axis viewing angles and analyze the time-resolved light curves and spectral diagnostics.

For the purpose of stable time-resolved spectral measurements, the displayed time axis is defined relative to the selected main photon-arrival window of each run rather than the full formal arrival-time extent of all escaped photons. Accordingly, the horizontal axis in Figure 9 should be interpreted as a relative analysis window for the selected escaped-photon sample, not as the full synthetic burst duration.

Figure 9 shows the results for $\theta_{obs} = 0°,\ 0.5°,\ 1°,$ and $2°$. The top panels present the normalized light curves (black) and $E_{pk}(t)$ (blue symbols; right axis). As the viewing angle moves off-axis, the pulse broadens and the peak time shifts systematically later within the displayed analysis window. This behavior is consistent with a reduced Doppler factor and the geometric broadening of EATSs.

At the current photon statistics, $E_{pk}(t)$ exhibits substantial scatter within individual time bins, making it difficult to infer a strictly monotonic temporal evolution for each viewing angle. Nevertheless, comparison across angles indicates that the characteristic time-integrated $E_{pk}$ decreases with increasing $\theta_{obs}$, consistent with Doppler deboosting and the weighted contribution from larger-angle emission.

The bottom panels show the effective low-energy index $\alpha_{eff}(t)$. For on-axis and mildly off-axis views (0° and 0.5°), $\alpha_{eff}$ remains positive for most of the pulse and varies relatively smoothly. At larger viewing angles, the temporal scatter increases and $\alpha_{eff}(t)$ can become negative in some intervals. However, within the present statistical precision, $\alpha_{eff}$ does not exhibit a strictly monotonic dependence on $\theta_{obs}$; instead, off-axis viewing primarily enhances the time-dependent structure and variability of the subpeak spectral shape.

We emphasize that the number of escaped photons within the acceptance cone decreases with viewing angle, amplifying statistical fluctuations in time-resolved spectral parameters. Therefore, interpretations of instantaneous indices in off-axis

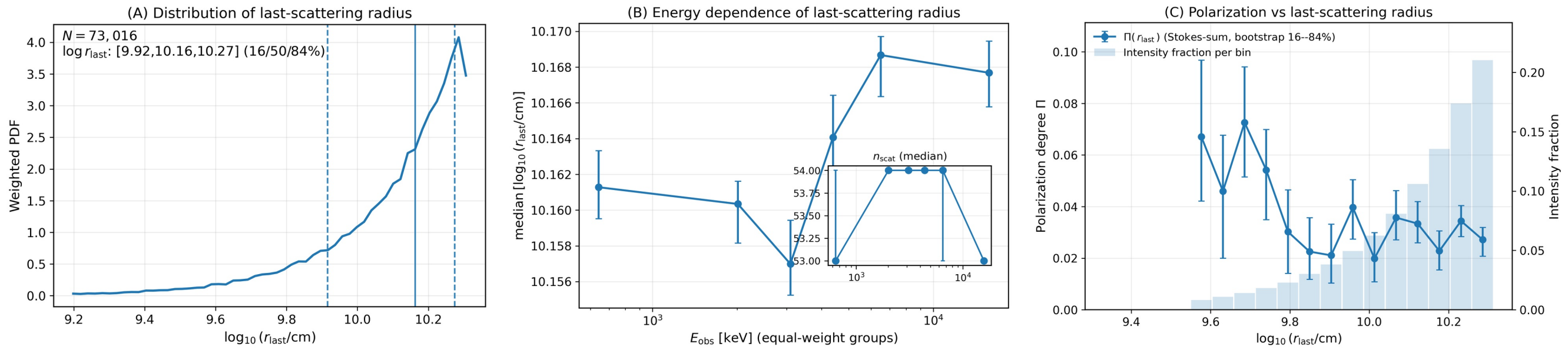

**Figure 10.** Statistical properties of photon-decoupling locations. (a) Weighted probability density of the last-scattering radius $r_{\rm last}$; vertical lines mark the weighted 16%, 50%, and 84% percentiles, showing that decoupling occurs over an extended radial range rather than a thin shell. (b) Weighted median $r_{\rm last}$ as a function of observed energy $E_{\rm obs}$ using equal-weight energy bins; error bars show the bootstrap 16%–84% intervals. The inset shows the corresponding median scattering number $n_{\rm scat}$. (c) Polarization degree $\Pi$ as a function of $r_{\rm last}$; error bars are bootstrap 16%–84% intervals, and the gray histogram indicates the relative intensity contribution from each radius bin.

cases must account for both statistical uncertainty and genuine physical evolution.

Overall, increasing $\theta_{\rm obs}$ broadens the pulse and delays the relative peak time within the selected analysis window, reduces the characteristic $E_{\rm pk}$, and enhances the amplitude and scatter of the time evolution of the low-energy spectral shape.

### *5.6. Decoupling Location Distribution*

To better understand the physical origin of the photospheric spectrum and polarization, we analyze the locations of the last-scattering events, i.e., where photons undergo their final interaction with the plasma before escape. In the Monte Carlo calculation, each escaped photon records its last-scattering radius $r_{\rm last}$, statistical weight, and Stokes parameters, enabling weighted statistics of the spatial structure of the decoupling region.

Figure 10(a) shows the weighted probability density distribution of $r_{\rm last}$ for all escaped photons. The last-scattering radii are not confined to a narrow geometric shell; instead, they span a finite but clearly broadened range of $\sim$0.3 dex. The weighted median and the 16%–84% percentiles are $\log_{10}(r_{\rm last}/{\rm cm}) \simeq 10.16^{+0.11}_{-0.25}$, demonstrating that decoupling occurs over an extended region rather than an idealized thin photospheric surface. This reflects the statistical nature of photon decoupling in a time-dependent flow, where the last-scattering location is shaped by both local dynamics and the radiative-transfer history.

Figure 10(b) shows the weighted median $r_{\rm last}$ as a function of observed energy $E_{\rm obs}$ using equal-weight energy grouping. The variation of $\log_{10}(r_{\rm last})$ with energy is limited and does not show a clear monotonic trend. At the current statistics, differences among energy groups are typically at the $\sim 10^{-2}$ dex level ($\lesssim$0.015 dex). The inset shows that the median scattering number $n_{\rm scat}$ varies similarly weakly across energy groups, implying that the mild energy dependence is unlikely to be driven by large differences in scattering counts and may instead be related to the spatiotemporal structure of the extended decoupling region.

Figure 10(c) shows the polarization degree $\Pi(r_{\rm last})$ as a function of decoupling radius. The polarization degree is computed by weighted summation of Stokes parameters within each radius bin, and uncertainties are estimated via bootstrap resampling. We find clear structural variations of $\Pi$ across the decoupling region, with typical values of a few percent and reaching $\sim$7% in some intervals. Notably, the radius range contributing most strongly to the total intensity (gray histogram) does not coincide with extreme polarization peaks, indicating that the observed net polarization is an aggregate of contributions from multiple radii rather than being dominated by a single shell.

In summary, in a time-dependent structured outflow the decoupling region is spatially extended, its energy dependence is mild and nonmonotonic, and the net polarization arises from a weighted superposition over an extended last-scattering region, rather than a single photospheric radius.

### *5.7. Connection to Published Spectropolarimetric GRBs*

Although the calculations presented here are not tuned to reproduce the absolute duration, luminosity, or peak-energy scale of any individual burst, it is still instructive to compare the main qualitative trends of the model with published spectropolarimetric GRB data. In particular, our results suggest that time-integrated spectral quantities may remain comparatively robust across different optical-depth treatments, whereas prompt polarization—especially its time dependence and energy dependence—is substantially more sensitive to the geometry of the last-scattering region. The three observational examples discussed below are summarized in Table 6.

A useful reference case is GRB 170114A, jointly observed by Fermi/Gamma-ray Burst Monitor and POLAR. Its prompt spectrum is characterized by fairly standard Band-like parameters, while the polarization analysis indicates a moderate but clearly time-dependent prompt polarization signal with polarization-angle evolution across the pulse (P. Veres 2017; J. M. Burgess et al. 2019). This is qualitatively consistent with one of the main conclusions of our dynamic calculations: even when the spectrum itself lies within an ordinary GRB range, the polarization can still retain strong sensitivity to the evolving decoupling geometry.

GRB 160802A provides a more extreme spectropolarimetric example. Its prompt emission shows a very high polarization degree in the 100–300 keV band, together with an unusually hard low-energy spectrum (V. Chand et al. 2018). While our present simulations do not attempt a dedicated fit to this event, such a combination is broadly compatible with a picture in which anisotropic photospheric decoupling and structured-jet geometry play an important role in shaping both the subpeak spectrum and the net polarization.

**Table 6**
Published Spectropolarimetric GRBs Used Here as Phenomenological Anchors for Comparison with the Model Trends

| GRB | Published Spectral Property | Published Polarization Property | Qualitative Connection to This Work |
|---|---|---|---|
| 170114A | Band-like prompt spectrum with $E_{pk} \sim 237$ keV, $\alpha \sim -0.84$, $\beta \sim -1.99$ | Moderate prompt polarization with clear time dependence and polarization-angle evolution | Supports the idea that polarization can be strongly geometry sensitive even when the time-integrated spectrum remains within a fairly standard GRB range |
| 160802A | Unusually hard low-energy prompt spectrum | Very high prompt polarization in the 100–300 keV band | Consistent with the possibility that structured decoupling geometry can jointly shape a hard subpeak spectrum and a large net polarization |
| 171010A | Bright prompt emission with characteristic spectral peak in the ∼100–200 keV range | Evidence for time-variable polarization and energy-dependent polarization behavior | Qualitatively consistent with a model in which different photon energies sample different parts of an extended last-scattering region |

A third relevant case is GRB 171010A, for which published analysis indicates time-variable prompt polarization and evidence that the polarization can differ across energy bands (V. Chand et al. 2019). This is of particular interest in the context of our model, because the calculations predict that polarization need not be a single energy-independent quantity: different photon energies can sample different parts of the extended last-scattering region and therefore carry different polarization signatures.

Taken together, these published bursts do not provide a one-to-one validation of the present model, and we do not claim such a correspondence here. However, they do show that several observational combinations emphasized by our calculations already occur in real data: moderate but evolving prompt polarization at otherwise standard spectral parameters, the coexistence of a hard subpeak spectrum and high polarization, and possible time- and energy-dependent polarization behavior. In this sense, the existing spectropolarimetric GRB sample is at least qualitatively compatible with the idea that polarization is a particularly sensitive diagnostic of structured, dissipative photospheric decoupling in relativistic jets.

## 6. Discussion and Conclusions

We have investigated time-resolved spectral diagnostics and polarization properties of photospheric emission from a structured jet by performing Monte Carlo radiative transfer on top of a two-dimensional, axisymmetric, time-dependent SRHD background.

Photon escape is determined using the residual optical depth along the instantaneous photon direction, $\tau_{out}(\hat{\Omega})$, while the SRHD background is treated as piecewise constant in time. This provides a time-dependent post-processing framework for studying how structured decoupling imprints itself on spectra and polarization.

A central aim of this work has been not only to compute synthetic observables but also to identify which of them are comparatively robust and which depend more strongly on the treatment of photon-decoupling geometry. To this end, we compared the full dynamic calculation with an axis-based reference treatment across the baseline model, the dissipation-window scan, and the pair-loading sequence and interpreted the differences using last-scattering statistics.

Our main conclusions are as follows:

1. *Baseline model (no dissipation, no pair loading).* The emission produces a single-peaked, nearly symmetric pulse, with only moderate evolution of $E_{pk}(t)$ and a narrow spectral width, $W = \log_{10}(E_{90}/E_{10}) \sim 0.30$–$0.35$, consistent with quasi-thermal photospheric emission. The similarity between different injection-depth realizations and between the dynamic and axis-based treatments shows that this baseline phenomenology is one of the least geometry-sensitive parts of the problem.
2. *Optical-depth window of subphotospheric dissipation.* The spectral outcome depends strongly on the depth at which dissipation operates. Shallow dissipation increases $E_{pk}$ and enhances nonthermal broadening, whereas deeper dissipation drives the spectrum toward stronger thermalization and a weaker high-energy tail. Intermediate dissipation windows are most favorable for substantial spectral broadening without full rethermalization. The dynamic–axis comparison shows that the qualitative ordering of this parameter space is robust, although the quantitative amplitudes depend on the full path-dependent reheating geometry. This optical-depth dependence is consistent with early dissipative-photosphere calculations, which showed that subphotospheric energy release modifies the emergent thermal component in different ways depending on whether dissipation occurs near the photosphere or at larger optical depth (A. Pe'er et al. 2006).
3. *Pair loading.* Increasing $Z_{\pm}$ generally raises $E_{pk}$ and softens the subpeak slope, while the high-energy tail is suppressed overall but can vary nonmonotonically. The spectral trends are qualitatively similar in the dynamic and axis-based calculations, but the polarization response differs much more strongly. In particular, the dynamic calculation predicts a substantially enhanced polarization degree at intermediate pair loading, whereas the axis-based reference does not. This indicates that polarization is more geometry sensitive than the corresponding time-integrated spectral diagnostics.
4. *Viewing-angle dependence.* As $\theta_{obs}$ increases, the pulse broadens and the peak time is delayed, consistent with Doppler deboosting and EATS broadening. The characteristic time-integrated $E_{pk}$ decreases with viewing angle, while time-resolved spectral parameters show larger scatter off-axis owing to both geometric mixing and reduced photon statistics.
5. *Extended decoupling region and polarization origin.* The last-scattering statistics show that photon decoupling occurs over an extended radial range rather than at a thin photospheric surface. The energy dependence of $r_{last}$ is mild and nonmonotonic, whereas the polarization degree varies structurally across the decoupling region. The observed net polarization is therefore produced by a weighted superposition over multiple radii and histories, naturally explaining why polarization responds more strongly than spectral quantities to the treatment of decoupling geometry.

6. *Geometry sensitivity hierarchy.* Taken together, our results indicate a clear hierarchy. Baseline pulse morphology, the narrow quasi-thermal width in the no-dissipation limit, the monotonic softening of $\alpha_{\rm eff}$ with increasing pair loading, and the overall ordering of the dissipation-window parameter space are comparatively robust. By contrast, the quantitative amplitude of dissipation-induced spectral changes, detailed time-resolved spectral evolution, and especially polarization are more sensitive to the full direction-dependent decoupling geometry. In this sense, spectral quantities are often sufficient for identifying first-order trends, whereas polarization provides a more discriminating probe of the geometric realism of the transfer calculation.

These results support a physically useful interpretation of structured photospheric emission in GRB jets.

We also emphasize that the present simulations are not intended as event-by-event fits to specific observed GRBs. The absolute pulse duration, peak-energy scale, and detailed temporal structure depend on the adopted jet power, background scale, injection conditions, and the simplified treatment of dissipation and adiabatic losses. The emphasis of this work is therefore on relative trends and diagnostic dependencies within a controlled parameter study. In this sense, the comparison with published spectropolarimetric GRBs should be understood as qualitative rather than one-to-one: the existing data are at least consistent with the picture that polarization remains sensitive to structured and anisotropic photospheric decoupling even when the time-integrated spectrum appears less distinctive.

The main limitations of this work are the parameterized treatment of subphotospheric dissipation and pair loading, the approximate post-processing treatment of adiabatic photon energy losses, the neglect of magnetic-field-related radiation processes, and the omission of radiative feedback on the hydrodynamic evolution. These approximations may affect the absolute normalization of quantities such as the pulse duration and peak-energy scale more strongly than the relative trends emphasized here. Future progress toward self-consistent radiation (magneto)hydrodynamic calculations with dissipation, pair creation, and radiative feedback will be essential for testing structured-jet photospheric models more rigorously and for improving joint inference of jet structure, dissipation depth, and pair content from combined spectral and polarization diagnostics.

## Acknowledgments

We thank the anonymous referee for helpful comments. This work is supported by the National Natural Science Foundation of China (grant No. 12573045) and the Jiangxi Provincial Natural Science Foundation (grant Nos. 20242BAB26012 and 20224ACB211001).

## Appendix
## Numerical Implementation Details of the SRHD Solver

This appendix summarizes numerical implementation details of the SRHD background used in this work, including the treatment of axisymmetric geometric source terms, the explicit expressions for the characteristic wave-speed bounds, the primitive-variable recovery procedure, and the floors/ceilings applied for robustness. Our implementation follows standard practices in relativistic hydrodynamics (J. M. Martí & E. Müller 1996, 1999; E. F. Toro 1999; R. J. LeVeque 2002), with choices tuned to prioritize smoothness and robustness of the background flow for subsequent radiative-transfer post-processing.

### A.1. Axisymmetric Geometric Source Terms and Near-axis Treatment

In axisymmetric cylindrical coordinates $(r,\ z)$, the SRHD equations in conservative form can be written as

$$\partial_t \boldsymbol{U} + \partial_r \boldsymbol{F}_r + \partial_z \boldsymbol{F}_z = \boldsymbol{S}_{\rm axi}, \tag{A1}$$

where $\boldsymbol{U} = (D,\ S_r,\ S_z,\ \tau)^{\top}$. In an ideal formulation, the radial flux divergence would be discretized as $\frac{1}{r}\partial_r(r\boldsymbol{F}_r)$. However, this form can become numerically fragile as $r \to 0$.

In the production version of the solver, the cylindrical geometry is therefore not evolved through an explicitly $r$-weighted finite-volume update. Instead, we adopt a Cartesian-like flux-divergence discretization in $(r,\ z)$ and add the geometric source contribution explicitly in the radial momentum equation. Specifically, we include the axisymmetric source term only in the $S_r$ equation,

$$S_{{\rm axi},\, S_r} = \frac{p}{r}, \tag{A2}$$

and in practice enforce a lower bound on $r$,

$$r \ \rightarrow\ \max(r,\ r_{\rm floor}), \qquad r_{\rm floor} = 10^{-12}, \tag{A3}$$

to avoid division by zero. This patch form was found to be numerically more robust near the symmetry axis in the present jet-background calculations, while preserving the large-scale jet structure needed for the subsequent radiative-transfer post-processing (J. M. Martí & E. Müller 1999; M. A. Aloy et al. 2000).

### A.2. Characteristic Wave-speed Bounds

The HLL flux requires estimates of the left and right signal-speed bounds, $s_L$ and $s_R$. These are obtained from the relativistic sound speed and the normal velocity. For an interface with normal direction $\hat{\boldsymbol{n}}$, define the normal velocity $v_n = \boldsymbol{v} \cdot \hat{\boldsymbol{n}}$. The characteristic speeds are then approximated by

$$\lambda_{\pm} = \frac{v_n \pm c_s}{1 \pm v_n c_s}, \tag{A4}$$

where the relativistic sound speed $c_s$ is given by Equation (5). After computing $\lambda_{\pm}^{(L)}$ and $\lambda_{\pm}^{(R)}$ on the left and right states, we set

$$s_L = \min(0,\ \lambda_-^{(L)},\ \lambda_-^{(R)}), \qquad s_R = \max(0,\ \lambda_+^{(L)},\ \lambda_+^{(R)}) \tag{A5}$$

as the wave-speed bounds for the HLL solver. This choice is known to be stable and broadly applicable in relativistic hydrodynamics (J. M. Martí & E. Müller 1996; E. F. Toro 1999).

### A.3. Primitive-variable Recovery: Pressure Bracketing and Bisection

Recovering primitive variables $(\rho,\ p,\ \boldsymbol{v})$ from the conserved state $\boldsymbol{U}$ can be particularly sensitive in the presence of high Lorentz factors and strong shocks. We adopt a cell-wise "pressure bracketing + bisection" strategy (J. M. Martí & E. Müller 1999).

Concretely, the recovery is reduced to solving a one-dimensional nonlinear root-finding problem for the pressure $p$. We perform a bisection search within a bracket $[p_{\min}, p_{\max}]$, where

$$p_{\min} = p_{\text{floor}}, \quad p_{\max} = \max(p_{\text{guess}},\ p_{\text{floor}}), \tag{A6}$$

and $p_{\text{guess}}$ is estimated either from the previous time step or from a local equilibrium-based approximation. We declare convergence when the relative change satisfies

$$\left| \frac{p^{(k+1)} - p^{(k)}}{p^{(k)}} \right| < \epsilon_{\text{tol}}, \qquad \epsilon_{\text{tol}} = 10^{-8}. \tag{A7}$$

Although slightly slower than closed-form or Newton-type approaches, this method is more robust in high-$\Gamma$ flows with strong shear and is well suited for producing smooth SRHD backgrounds for radiative post-processing.

### *A.4. Floors and Ceilings*

To prevent numerical vacuum states and other unphysical conditions, we enforce the following floors on primitive variables at each time step:

$$\rho \geqslant \rho_{\text{floor}}, \qquad p \geqslant p_{\text{floor}}, \tag{A8}$$

where $\rho_{\text{floor}}$ and $p_{\text{floor}}$ are chosen based on the minimum physical scales of the ambient medium. We also impose an upper bound on the sound speed squared,

$$c_s^2 \leqslant c_{s,\max}^2 < 1, \tag{A9}$$

to ensure physical characteristic speeds and numerical stability.

We emphasize that these floors and ceilings are activated only in extremely low density regions or near numerical boundaries and do not qualitatively affect the jet core structure or the subsequent photospheric radiative-transfer results. Their primary role is to maintain a smooth and usable background throughout long-term evolution.

### *A.5. Representative Compression Structure in the SRHD Background*

To document the hydrodynamic structure of the SRHD background more explicitly, Figure A1 shows transverse cuts through the compressed jet-head region at two representative output times. The cuts are taken at fixed axial positions near the head/interface region and display the radial profiles of density, pressure, and Lorentz factor. These diagnostics reveal strong transverse gradients and compressed structure associated with the jet core, the jet–cocoon interface, and the surrounding shocked flow. Their purpose here is diagnostic: they demonstrate that the background used for the radiative-transfer post-processing contains resolved large-scale hydrodynamic structure rather than a featureless smooth outflow.

In particular, the Lorentz factor profiles show a sharp transition from the relativistic jet core to the surrounding slower material, while the density and pressure profiles exhibit corresponding structured variations across the same radial range. The persistence of these features at multiple output times indicates that the SRHD background retains an evolving

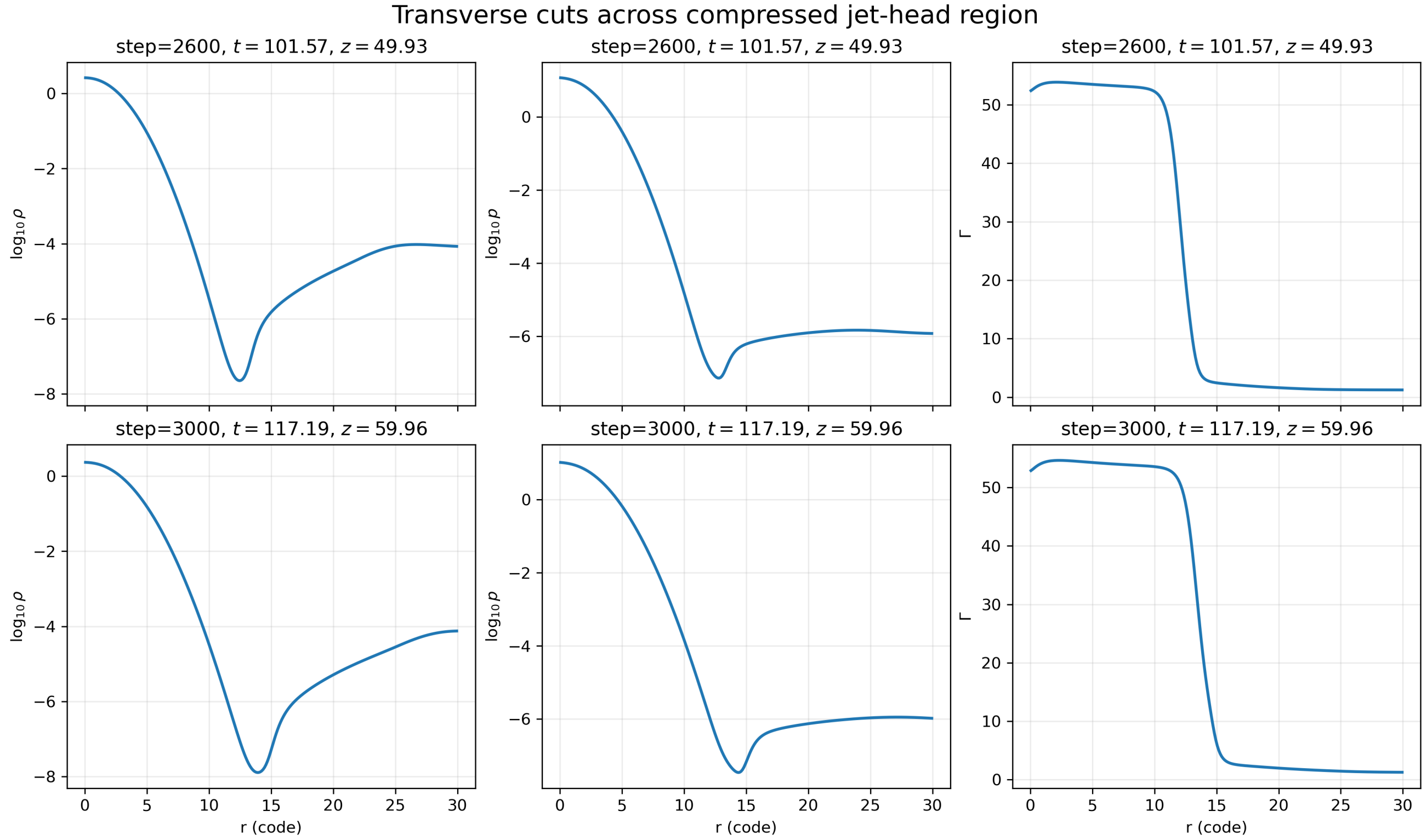

**Figure A1.** Transverse cuts across the compressed jet-head region of the SRHD background at two representative output times. The upper and lower rows correspond to steps 2600 and 3000, respectively, with the cuts taken at fixed axial positions near the jet-head region. From left to right, the panels show $\log_{10}\rho(r)$, $\log_{10}p(r)$, and $\Gamma(r)$. All three quantities exhibit strong transverse gradients across the jet-head/jet–cocoon interface, including a sharp drop in Lorentz factor and pronounced structure in density and pressure. These profiles are included to document the large-scale compression structure present in the SRHD background used for the Monte Carlo radiative-transfer calculations. The present work does not attempt to compute shock-powered radiation self-consistently; instead, radiative dissipation is modeled through the parameterized optical-depth window described in the main text

but well-resolved compression/interface structure suitable for Monte Carlo post-processing.

We emphasize, however, that the present paper does not model shock-powered radiation self-consistently. The Monte Carlo calculations employ a parameterized subphotospheric dissipation prescription, introduced through an optical-depth window, in order to study how spectra and polarization respond to dissipation depth, pair loading, and geometry in a time-dependent structured background. Accordingly, the role of Figure A1 is to document the hydrodynamic background and its compression structure, not to identify a one-to-one mapping between individual hydrodynamic compression features and radiative emission sites.

## ORCID iDs

Qingwen Tang https://orcid.org/0000-0001-7471-8451